\documentclass{emulateapj}

\newcommand{\U}[1]{\ensuremath{\mathrm{~#1}}}
\newcommand{\Myr}{\U{Myr}}
\newcommand{\Gyr}{\U{Gyr}}
\newcommand{\kpc}{\U{kpc}}
\newcommand{\msun}{\U{M}_{\odot}}

\begin{document}
\title{Fully compressive tides in galaxy mergers}

\author{F.~Renaud\altaffilmark{1,2}, C.~M.~Boily\altaffilmark{1}, T.~Naab\altaffilmark{3} and Ch.~Theis\altaffilmark{2}}
\altaffiltext{1}{Observatoire astronomique and CNRS UMR 7550, Universit\'e de Strasbourg, 11 rue de l'Universit\'e, F-67000 Strasbourg, France}
\altaffiltext{2}{Institut f\"ur Astronomie der Univ. Wien, T\"urkenschanzstr. 17, A-1180 Vienna, Austria}
\altaffiltext{3}{University Observatory, Scheinerstr. 1, D-81679 Munich, Germany}
\email{renaud@astro.u-strasbg.fr}

\begin{abstract}

The disruptive effect of galactic tides is a textbook example of gravitational dynamics. However, depending on the shape of the potential, tides can also become \emph{fully} compressive. When that is the case, they might trigger or strengthen the formation of galactic substructures (star clusters, tidal dwarf galaxies), instead of destroying them. We perform $N$-body simulations of interacting galaxies to quantify this effect. We demonstrate that tidal compression occurs repeatedly during a galaxy merger, independently of the specific choice of parameterization. With a model tailored to the Antennae galaxies, we show that the distribution of compressive tides matches the locations and timescales of observed substructures. After extending our study to a broad range of parameters, we conclude that neither the importance of the compressive tides ($\approx 15\%$ of the stellar mass) nor their duration ($\sim 10^7 \U{yr}$) are strongly affected by changes in the progenitors' configurations and orbits. Moreover, we show that individual clumps of matter can enter compressive regions several times in the course of a simulation. We speculate that this may spawn multiple star formation episodes in some star clusters, through e.g., enhanced gas retention.

\end{abstract}

\section{Introduction}

That the violent collision at the origin of a galaxy merger acts to develop strong tidal features is a long-acknowledged fact of its evolution in time, as encapsulated in the Toomre sequence \citep[see several illustrations in][]{Sandage1961, Toomre1972, Toomre1977, Hibbard1996, Laine2003, Elmegreen2007}. The burst of star formation triggered during such an event is a catalyst for the evolution of entire galactic stellar populations \citep{Barton2000, Bridge2007, Li2008}. This process would be even more significant at high redshift ($z \gtrsim 1$, \citealt{Madau1996}), through either repeated accretion during the formation of a major galaxy, or enhanced merger rates driven by the formation of cosmological large-scale structures. However, the lack of resolution at these redshifts is an intrinsic difficulty when trying to identify stellar associations down to a few thousand solar masses, of a size spanning a few parsecs. At today's resolution $\sim 0.01\arcsec$, a full coverage of all modes of star formation triggered during a merger limits surveys to the low redshift, local Universe.

Over the past decade, high-resolution HST observations of the proto-typical merger NGC 4038/39 (the Antennae) have revealed a surprisingly high number of dense, massive young clusters of stars in its central region \citep[][]{Whitmore1995, Meurer1995, Mengel2005, Whitmore2007}. A key question which emerges from such studies pertains to the likelihood that stellar associations will survive in the rapidly-varying background potential of the merger, once they have outlasted the early phase of internal evolution \citep{Elmegreen1997, Bastian2006, Gilbert2007}. Based on samples of some $10^3$ Antennae clusters, \citet{Fall2005} derived a feature-less power-law age distribution ($dN/d\tau \propto \tau^{-1}$) with a median value of $\sim 10^7 \U{yr}$, a trend they argued to be independent of mass. Their analysis may be interpreted as implying that young clusters dissolve rapidly into their gas cloud nursery, prior to reaching virial equilibrium \citep{Whitmore2007,deGrijs2007}. \citet{Gieles2009} has shown that mass-dependent models of cluster evolution leading to dissolution would inevitably introduce a feature in the age distribution of Antennae clusters, lending support to the conclusions of \citet{Fall2005}. Since cluster dissolution in equilibrium galaxies invariably lead to Gaussian distributions \citep{Vesperini1997, Baumgardt2003}, one is drawn to ask whether the power-law properties of young cluster distributions in rapidly evolving, out-of-equilibrium mergers are shaped mainly by gravitational effects, or hydrodynamics, or both. Clearly the evolution of the large-scale gravitational potential imprints the smaller scales of star formation controlled by the hydrodynamics of accretion seeds \citep{Klessen1998, MacLow2004}. Knowing how to connect the two would be a major step toward a global understanding of the evolution of stellar populations in mergers.

One way of addressing this problem consists in recreating observations numerically. Simulations of galaxy mergers present much more complicated kinematics that increase the complexity level of star formation modeling when compared, say, to galaxies taken in isolation. To resolve the many star-forming regions in a galaxy merger, several studies rely on smoothed particle hydrodynamics \citep[SPH,][]{Gingold1977} simulations, assuming a star formation recipe mainly based on a local density or pressure threshold \citep{Mihos1996, Bekki2002a, Bekki2002b, Springel2003, Li2004, Robertson2008, Karl2008, Johansson2009}. Other approaches like sticky particles \citep[e.g.][]{Bournaud2008} or adaptive mesh refinement \citep{Kim2009} are also employed to investigate the formation of substructures (star clusters or tidal dwarf galaxies, TDGs) at high resolution. \citet{Barnes2004} compared the common SPH star formation rule with a new prescription based on the energy dissipation through shocks, with an application to NGC 4676 (the Mice). He showed that to gauge correctly the radiative energy from shocks and to recover a more realistic star-formation rate, it is necessary to take into account the divergence of the background velocity field, a strong hint that large-scale motion bears down on the small-scale physics of star formation. \citet{Dobbs2009} recently showed that the surface density of neither the total gas (\ion{H}{1} + H$_2$) nor the warm gas only (\ion{H}{1}) were good tracers for the local star formation rate, considering an isolated disk. They also underlined that a strong spiral shock would increase the local density, but the velocity dispersion too, which preempts a one-to-one relation between star formation (averaged over the whole disk) and shocks. These results suggest to consider (on top of a density contrast) dynamical properties such as the convergence of the velocity field, either through hydrodynamical shock dissipation or through a background gravitational potential. On a smaller scale, several studies starting with \citet{Ebert1955} and \citet{Bonnor1956}, investigated the hydrodynamics of gaseous systems like giant molecular clouds confined by an external pressure \citep[see e.g.][]{Kumai1986, Elmegreen1989, Harris1994, Elmegreen1997}. They showed that such an external effect could trigger the collapse of the cloud, so linking galactic-scales hydrodynamics and cluster-size phenomena. Despite this impressive body of work, the high-resolution needed to link gravitational galactic dynamics to the very small scales of star formation is still out-of-reach of current 3D numerical models.

A possible way out of this dilemma would be to impose the right boundary conditions on a small volume within which high-resolution hydro codes may resolve the physics of star formation. This paper takes a step in that direction and investigates how these boundary conditions might evolve as a function of time. \citet{Renaud2008} already explored the evolution of the tidal field during a merger event, with a model tailored to the Antennae galaxies. They showed that \emph{fully} compressive tides (see Section~\ref{sec:tidaltensor}) develop during the interaction of two disk galaxies and last for few $\times 10^7 \U{yr}$. They concluded that such a tidal mode could play an important role in triggering the formation of star clusters, or at least prevent them from destruction. Here, we detail further and extend these results through a larger sample of progenitor galaxies and merger orbits. It is hoped that this more general yet precise description of the tidal field will in turn allow more realistic boundary conditions for volumes where star formation takes place. In Section 2, we present analytically how compressive tides may develop when two potentials overlap, as in the case of a major galaxy merger. Section 3 describes the application to $N$-body simulations and the limitations of the method. Section 4 details the Antennae reference model and Section 5 makes direct comparisons between the parameter survey and this reference. We close with a discussion on the link between the physical process described and the observational facts.

\section{Gravitational tidal tensor}
\label{sec:tidaltensor}

The tidal field describes how the internal dynamics of an extended body, e.g. a star cluster, is stirred by the influence of an external source of gravitation, say the host galaxy. In this case, the differential force due to the galaxy which exists between any two points within the cluster, and separated by $\ell$ is given roughly by $\delta F = \ell\ T$ where $T$ is the tidal tensor of components
\begin{equation}
\label{eqn:tidal_tensor}
T^{ij} \equiv -\partial_i \left(\partial_j \phi \right),
\end{equation}
which are minus the second derivatives of the gravitational potential $\phi$ along the $j$-th and $i$-th spatial axes. Because the tidal tensor is symmetric and real-valued, it can be set in orthogonal form. In this representation, the eigenvalues $\lambda$ gauge the strength of the tide along the corresponding eigenvectors. According to the sign of an eigenvalue, one may characterize the tide down the associated eigenvector to be compressive ($\lambda < 0$) or extensive ($\lambda > 0$).

The trace of $T$ reads
\begin{equation}
\label{eqn:poisson}
\sum_{i=1}^3 -\partial_i \left(\partial_i \phi \right) = -\nabla^2 \phi = -4\pi G \rho,
\end{equation}
where $\rho$ is the density and the last steps follows from Poisson's law. Since the density is a positive-definite quantity at every point in space, it is impossible to compute three positive eigenvalues simultaneously. Hence, the number of axes along which the eigenvalue may be positive ranges from zero to two. Cases for which one or two eigenvalues are positive occur in well-known configurations \citep[Earth-Moon interaction, galactic or cluster tides, spiral waves, see e.g.][]{Valluri1993, Dekel2003, Gieles2007}. However, the case where all three eigenvalues are negative has been somewhat overlooked in the literature. When that occurs, the tidal force field converges to a common point from all directions, and so the tides are \emph{fully} compressive with respect to that point. Hereafter, we will refer to such a situation as a \emph{compressive mode} of the tidal field.

From a mathematical standpoint, one may remark that a compressive mode only exists in the convex intervals of the twice-differentiable gravitational potential\footnote{The tidal tensor is the Hessian matrix of the potential and thus, vanishes at the inflection points.}. A simple analytical exercise demonstrates indeed that cored potentials, e.g. \citet{Plummer1911} and \citet[][for $0 \le \gamma < 1$]{Dehnen1993}, show a fully compressive region near their center of mass, while profiles with a cusp, e.g. point-mass, \citet{Hernquist1990}, \citet[][for $1 \le \gamma < 3$]{Dehnen1993} or NFW \citep{Navarro1997}, have extensive tides along at least one axis everywhere in space. (Table~\ref{tab:tensors} lists the analytical expressions of the cases we considered.)

\begin{deluxetable*}{ccc}
\tablecaption{Tidal tensors of typical density profiles\label{tab:tensors}} 
\tablehead{\colhead{Profile} & \colhead{$\phi(x = r/r_0)$} & \colhead{$T^{xx}(x=r/r_0, y=0, z=0)$}} 
\startdata
Point mass & $\displaystyle{-\frac{GM}{x}}$ & $\displaystyle{\frac{2GM}{x^3}}$ \\
\\
Plummer & $\displaystyle{-\frac{GM}{r_0\sqrt{1+x^2}}}$ & $\displaystyle{-GM\frac{1-2x^2}{r_0^3 \left( 1 + x^2\right)^{5/2}}}$ \\
\\
Hernquist & $\displaystyle{-\frac{GM}{r_0 (1+x)}}$ & $\displaystyle{\frac{2GM}{r_0^3 \left(1+x\right)^3}}$ \\
\\
Dehnen\tablenotemark{*} & $\displaystyle{-\frac{GM}{r_0(2-\gamma)}\left[ 1 - \left( \frac{x}{1+x}\right)^{2-\gamma} \right]}$ & $\displaystyle{GM\frac{2x+\gamma-1}{r_0^3 \ x^{\gamma} \ (1+x)^{4-\gamma}}}$\\
\\
NFW & $\displaystyle{- \frac{G \rho_0 \ r_0^2}{x} \ln{\left(1+x\right)}}$ & $\displaystyle{\frac{G \rho_0}{x^3} \left[ 2\ln{\left(1+x\right)} - \frac{3x^2+2x}{\left(1+x£\right)^2}\right]}$
\enddata
\tablenotetext{*}{For $\gamma = 2$, the potential is given by $GM/r_0 \ \ln{\left[x / (1+x)\right]}$ but the expression of the tidal tensor remains the same.}
\end{deluxetable*}

\begin{figure}
\plotone{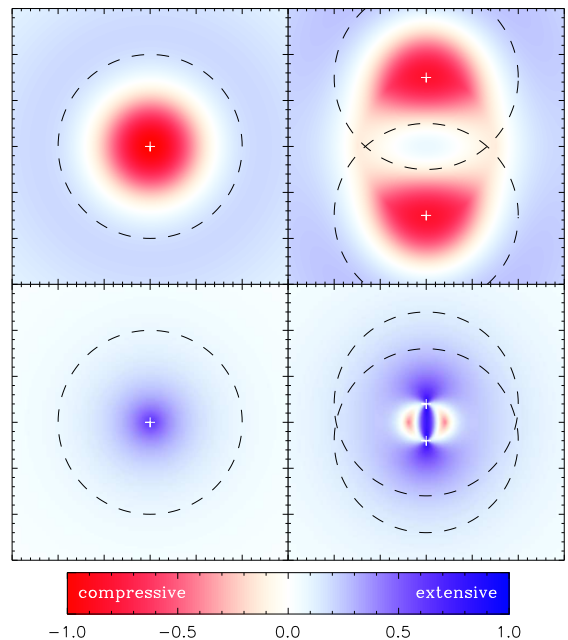}
\caption{Maximum value of the tidal tensor. The Plummer potential (top) shows an intrinsic compressive region (red) in isolation (top left) as well as for an overlap of two similar potentials (top right) separated by $1.5\ r_0$ (where $r_0$ is the core radius). For the Hernquist potential (bottom), the overlap of the two wells (bottom right) separated by $0.4\ r_0$ also presents a compressive region. However the isolated case (bottom left) is completely extensive. The dashed circles represent the characteristic radii $r_0$.}
\label{fig:analytical}
\end{figure}

The overlay of two identical density profiles (cuspy or not), one shifted with respect to the other, can modify the potential well such that compressive modes now appear where none was present in isolation. In Fig.~\ref{fig:analytical}, we show the examples of Plummer (top) and Hernquist (bottom) potentials, in isolation (left) and when two identical profiles overlap, one shifted along the vertical axis (right). The color code on that figures gives blue for areas where at least one tidal component is extensive, and red for a fully compressive mode. The Plummer potential in isolation has a large volume about the center where the tidal field shows a compressive mode; this situation does not occur in the case of a Hernquist profile. By contrast, compressive modes exist in both cases where two potentials overlap (right-hand frames). This demonstrates that even for the case when an isolated galaxy hosts no compressive tides anywhere, such tides may still develop widely as two galaxies come closer and merge.

The situation depicted in Fig.~\ref{fig:analytical} will occur naturally in the course of a galaxy merger. When a progenitor galaxy (sum of a disk, bulge and dark matter halo) comes close enough to a second, similar galaxy, local cores in the global potential may appear and lead to compressive modes of tide that may last for a significant period of time. As a side issue, it follows that a compound model of a progenitor galaxy may itself display regions where tides are compressive although each component, taken individually, may not. This phenomenon would in turn affect the matter (stars, clusters, interstellar medium) inside these areas and possibly alter their dynamical and chemical properties (e.g., through enhanced retention of metals). However, the very existence of such tidal modes will be impossible to predict in a realistic simulation of a galaxy merger with analytic tools only, owing to the irregular morphology of the system.

\section{Numerical method}
While simple density profiles allow an analytical examination of the tidal field, the same approach applied to mergers is clearly out of reach. To overcome this, we set up three-dimensional $N$-body simulations to explore the time-evolution of the gravitational potential and its associated tidal field. This section gives details of the numerics used in this work.

\subsection{Method}

For simplicity, we used purely gravitational models (no gas). This approach holds good as long as the gas represents a small fraction of the total mass of the galaxy, which is usually the case in the local Universe. Therefore neglecting the gas component should not affect significantly the shape of the gravitational potential and the characteristics of its tidal tensor. The numerical integration of orbits was performed with the public version of Walter Dehnen's code {\tt gyrfalcON} with a compact kernel K0 \citep[see][]{Dehnen2002}. The mass- and length resolutions were linked to typical globular cluster scales (i.e. $10^5 \msun$ and $40 \U{pc}$). All the particles have the same mass to reduce two-body relaxation.

In our compound $N$-body galaxy models (disk + bulge + dark matter halo), we focused on disk particles when computing the tidal tensor, since, being dynamically cold, they give rise to large-scale arms, bridges, and other transient features. To compute the tensor $T$, first a mass-less particle is placed on each of the six faces of a cube centered on a given disk-particle coordinates. In a second step, the force field is computed at the position of each pseudo-particle using the {\tt getgravity} tool \citep{Dehnen2002}. Note that the gravity of the disk particle sitting at the centre of the cube is subtracted explicitly. Finally, we implemented a first-order-accurate differencing scheme to compute the gradients of the forces by pairing the pseudo-particles as needed to derive the six independent components of the symmetric tidal tensor. The eigenvalues and eigenvectors were then computed efficiently using the method of \citet{Kopp2008}. The sign of the largest eigenvalue determines whether the central disk particle sits in a fully compressive volume, or not.

In order to obtain statistically significant results, we applied the above procedure to all the disk particles of every simulation snapshot (each separated by a time interval of $\simeq 2.5 \Myr$). In that fashion we retrieved the evolution of the maximum eigenvalue to trace the history of the tidal field of disk particles, a binary test on the character of the tide (compressive or extensive) experienced by a particle throughout the merger simulation.

\subsection{Tests}
\label{sec:tests}

Using the analytical Plummer model (see Table~\ref{tab:tensors}) as a test case, we computed the gradients of the force (i.e. the tidal tensor) using a first-order-accurate finite difference scheme. The size of the cube $\delta$ defines a length for the differentiation scheme which should be chosen to minimize errors. Large values of $\delta$ will fail to resolve fine features in the potential, while very small values are prohibited when round-off errors kick in. In practice, however, errors attributable to the difference scheme accounted for less than $15 \%$ of the total uncertainties: they have been omitted from the remainder of the discussion, for the sake of clarity.

The main source of errors comes from the $N$-body rendition of the potential. When integrating the orbits of an $N$-body model, a softening length $\epsilon$ is often invoked to suppress small-scale errors in the force field. This length also sets a resolution scale for all quantities bearing on the dynamics. Thus, one must choose $\delta > \epsilon$ to compute the tidal tensor coherently with the resolution of the simulation. The more tricky issue of finding a suitable upper bound for $\delta$ for a given $\epsilon$ remains.

\begin{figure}
\plotone{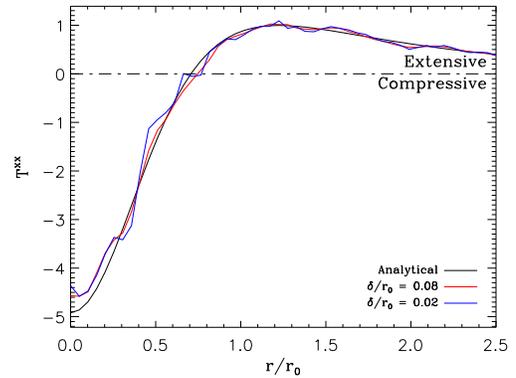}
\caption{Maximum eigenvalue of the tidal tensor of a Plummer sphere along one radius. A cube size of $\delta /r_0 = 0.08$ (with $r_0$, the core radius) minimizes the dispersion ($\sigma \simeq 0.13$) relative to the analytical solution (see Table~\ref{tab:tensors}), while smaller and larger values lead to less precise fits (e.g. a size $\delta /r_0 = 0.02$ shows $\sigma \simeq 0.19$). Small cubes increase the noise while larger ones do not match the analytical solution. For a Plummer distribution, the compressive region is a sphere of radius $r_0/\sqrt{2}$, which encloses $\simeq 19\%$ of the total mass.}
\label{fig:testplummer}
\end{figure}

To address this question, a series of tests were performed with an $N = 5 \times 10^5$ bodies realization of a Plummer sphere. Eigenvalues and eigenvectors were computed throughout the model with the numerical scheme outlined above and for various values of $\delta$. The best results (Fig.~\ref{fig:testplummer}) compared to the analytical values had $\delta \approx 0.08\ r_0 \approx 0.05$, where $r_0$ is the Plummer core radius.

This value of $\delta$ may be compared to the softening $\epsilon$ of the force field \citep{Merritt1996, Dehnen2001} which has been set to $0.01$ for the choice of $N$ adopted here. Therefore, an empirical relation suggests itself $\delta \approx 5 \epsilon$.

To determine the scope of this relation, we repeated the calculations outlined here to several compound models of the Antennae galaxies, marking a clear break with the spherically symmetric Plummer potential, with values of $N$ ranging from 350,000 to $1.4 \times 10^6$. We found that the relation $\delta = 5\epsilon$ holds well over this range of particle number. It was therefore applied to all simulations presented in the following parameter survey.

\section{The Antennae as a reference}
\label{sec:antennae}

The starting point of our survey is the model of the Antennae galaxies presented in \citet{Renaud2008}. It consists of two self-gravitating equal-mass progenitors. They have been set up with {\tt magalie} \citep{Boily2001} which gathers an exponential disk, a \citet{Hernquist1990} bulge and an isothermal dark matter halo. The mass ratios and length-scales of each component have been set to get two S0 galaxies. Initial parameters are described in Table~\ref{tab:numsetup}.

\begin{deluxetable}{lcc}
\tablecaption{Parameters of the reference Antennae model\label{tab:numsetup}}
\tablehead{\colhead{Parameter} & \colhead{NGC 4038} & \colhead{NGC 4039}}
\startdata
\multicolumn{3}{l}{Numbers of particles}\\
$N_{\mathrm{disk}}$ & $2 \times 10^5$ & $2 \times 10^5$ \\
$N_{\mathrm{bulge}}$ & $1 \times 10^5$ & $1 \times 10^5$ \\
$N_{\mathrm{halo}}$ & $4 \times 10^5$ & $4 \times 10^5$ \\
\hline
\multicolumn{3}{l}{Scale-lengths}\\
$h_{\mathrm{disk, vertical}}$ & $0.2$ & $0.2$ \\
$r_{\mathrm{disk}}$ & $1.0$ & $1.0$ \\
$r_{\mathrm{bulge}}$ & $1.0$ & $1.0$ \\
$r_{\mathrm{halo}}$ & $7.0$ & $7.0$ \\
\hline
\multicolumn{3}{l}{Cut-off radii}\\
$c_{\mathrm{disk, vertical}}$ & $2.0$ & $2.0$ \\
$c_{\mathrm{disk}}$ & $5.0$ & $3.0$ \\
$c_{\mathrm{bulge}}$ & $1.0$ & $1.0$ \\
$c_{\mathrm{halo}}$ & $7.0$ & $7.0$ \\
\hline
\multicolumn{3}{l}{Mass ratios}\\
$m_{\mathrm{disk}}$ & $1.0$ & $1.0$ \\
$m_{\mathrm{bulge}}$ & $1.0$ & $1.0$ \\
$m_{\mathrm{halo}}$ & $5.0$ & $5.0$ \\
\hline
\multicolumn{3}{l}{Toomre parameters}\\
$Q$ & $1.5$ & $1.5$ \\
\hline
\multicolumn{3}{l}{Disks inclinations}\\
$\theta_x$ & $60^{\circ}$ & $-60^{\circ}$ \\
$\theta_y$ & $0^{\circ}$ & $30^{\circ}$ \\
\hline
\multicolumn{3}{l}{Initial coordinates}\\
$(x,y,z)$ & $(6.0, 6.0, 0.0)$ & $(-6.0, -6.0, 0.0)$ \\
$(v_x,v_y,v_z)$ & $(-0.5, -0.25, 0.0)$ & $(0.5, 0.25, 0.0)$
\enddata
\tablecomments{The parameters are given in numerical units ($G=1$) with the conversion factors $4.4 \kpc$, $1.9 \times 10^2 \U{km.s^{-1}}$ and $3.6 \times 10^{10} \msun$ for the length, velocity and mass respectively. The non-dimensionalization is based on the spatial extension of the tidal tails and the peak in the velocity field measured by \citet{Hibbard2001}, adopting a distance to the Antennae of $19.2 \U{Mpc}$. The Keplerian equivalent orbit has an eccentricity of 0.96, a distance to pericenter of 1.31 ($= 5.78 \kpc$) and a semi major axis of 17.74 ($= 78.06 \kpc$).}
\end{deluxetable}

A total of 1,400,000 particles have been used. As mentioned before, our resolution matches the typical cluster scale ($\epsilon \simeq 40 \U{pc}$ and $10^5 \msun$). The progenitor galaxies have been set on a bound orbit with an eccentricity $e \approx 0.96$ and an initial separation of $75 \kpc$. According to the relation $\delta = 5 \epsilon$ (see Section~\ref{sec:tests}), we set the size of our cubes to $220 \U{pc}$, i.e. well beyond the typical half-mass radius of a star cluster.

\subsection{Time evolution}
\begin{figure}
\plotone{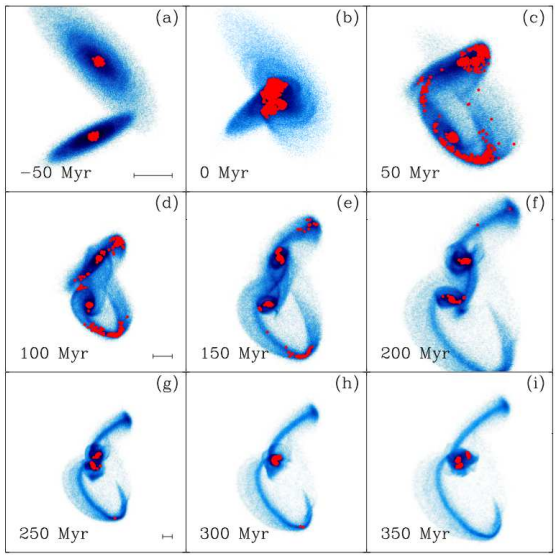}
\caption{Morphology of the disks of the Antennae model every $50 \Myr$. On each row, the black line represents $10 \kpc$. $t=0$ (b) corresponds to the first pericenter. The tidal bridges and tails are created in the first $10^7 \U{yr}$ after the first passage, while the galaxies move apart. The best match with observations is obtained at $t\approx 300 \Myr$ (h). Red dots denote the particles standing in fully compressive tidal mode.}
\label{fig:morphology}
\end{figure}

Fig.~\ref{fig:morphology} tells the history of the interaction. Most of the tidal features originate from the disks during the first pericenter passage. The two long tails expand as the galaxies move apart from each other. The central regions are strongly mixed during the second passage and brake significantly, which leads to an oscillatory phase and finally a rapid merger. The best match with observations (morphology and kinematics) occurs at the beginning of the second passage (Fig.~\ref{fig:morphology}, panel h), some $300 \Myr$ after the first one.

The evolution of the compressive tidal mode (in term of intensity and mass fraction) clearly shows the same steps. Fully compressive regions are displayed as red dots in Fig.~\ref{fig:morphology}. Initially, they are located in the intrinsically compressive regions at the very center of the galaxies (panel a) but, as the potentials overlap, they quickly propagate to the tidal arms and bridges (panels c, d, e). At $t \approx 150 \Myr$ (panel e), a compressive spiral structure is visible in the central part of the disks. Note on this panel a large ($\sim 20 \kpc$) compressive zone exists near the tip of the southern tail, matching the positions of the tidal dwarf galaxy candidates identified by several authors \citep{Schweizer1978,Mirabel1992}. We also note that age estimates of $\sim 120 \Myr$ from \citet{Hibbard2005} correspond well to our results.

During the second passage and the merger phase (panels g, h, i), the fast evolution of the mass distribution mixes the material in the central parts, which creates patterns of compressive mode, especially in the overlap region and around the two nuclei. Using this configuration of the potential, \citet[][Fig.~2]{Renaud2008} derived a map of the compressive regions. They used a 3D grid-based method applied on the potential given by the $N$-body simulation to get rid of projection effects. The good match between this map and the candidate clusters identified by \citet{Mengel2005}, in addition to the TDGs as discussed above, is a strong hint that the compressive tides are to be linked to the formation of substructures.

\begin{figure}
\plotone{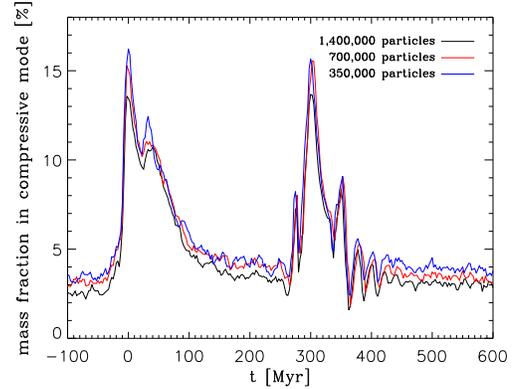}
\caption{Evolution of the mass fraction of the disks of the Antennae in compressive mode. The peaks denote the main steps of the merger such as the pericenter passages (at $t = 0$ and $t = 300 \Myr$). Lower resolutions give similar results (with deviations from the 1,400,000 resolution of $\sigma = 0.79$ and $\sigma = 1.0$ for 700,000 and 350,000 particles respectively).}
\label{fig:histo_resolution}
\end{figure}

Fig.~\ref{fig:histo_resolution} traces the evolution of the mass fraction in compressive mode along time. Results for lower resolutions (factors 1/2 and 1/4) are also plotted to demonstrate that the derived quantities are independent of the number of particles used to represent the galaxies. The main peaks correspond to the two pericenter passages and the smaller spikes are linked to the formation of smaller structures such as bridges.

In the background of all these steps, an intrinsic subset ($\sim 3\%$) of the stellar mass stays in compressive mode in the galaxies' centers (within a radius of $\sim 1.3 \kpc$, when the disks are at rest). This offset is due to the initial mass distribution and thus, will change with e.g. the dark matter halo shape. Because of orbital mixing, some particles will stay in this region for a very long time, but others will be replaced, keeping this fraction almost exactly constant (deviation of $0.2 \%$ over $12 \Gyr$).

\subsection{Characteristic times}
\label{sec:charcteristic_times}
   
For an isolated progenitor, the most bound particles stand in the central part of the potential well and thus, in the intrinsic compressive region. They will stay in this region for an arbitrary long time provided their energy is smaller than the energy level at the boundary of the compressive region. The fraction in this regime (constantly compressive) represents $\sim 1.1\%$ of the total disk mass\footnote{Some $\sim 1\%$ of these particles will leave the compressive region, over $1 \Gyr$, because of two-body relaxation.}. In the following, we subtracted these objects from the statistics, and concentrated on the particles entering compressive modes. The others $\sim 2\%$ in the compressive region (which gathers $2.0\% + 1.1\% \simeq 3.0\%$) enter and leave the central part because of the larger angular momentum of their highly eccentric orbits. Therefore, the duration of the compressive mode for individual particles (in an isolated galaxy) can be studied by considering their energy and angular momentum and deriving how much time they will spend inside the central compressive region. A good knowledge of these statistics allows to characterize the effect of the intrinsic tidal field, regarding to the influence of the merger.

At this point, we introduce two timescales. The \emph{longest uninterrupted sequence} (\emph{lus}) represents the duration of the longest continuously compressive episode, while the \emph{total time} (\emph{tt}) is the count of all compressive snapshots which might be separated by periods of extensive tides. As a particle would probably undergo a series of on/off compressive episodes, the \emph{tt} should be larger than, or equal to, the \emph{lus}. The limit is reached for the $1.1\%$ constantly compressive particles whose \emph{lus} equals \emph{tt}. A second family of particles counts those which orbits allow to enter the compressive region for a short timescale. These particles present a medium energy, a low angular momentum and thus an intermediate valued \emph{lus}. The last part gathers the high energy and high angular momentum particles that merely never enter into the compressive region. Their \emph{lus} should be extremely small.

The value of the angular momentum at the edge of the compressive region sets the upper limit of the particles which could experience compressive tides, during at least one snapshot ($2.5 \Myr$). Particles with a higher angular momentum would not enter the central region, while those with a smaller value could get into it. The last represents $7.7 \%$ of the disk mass. Therefore, because of orbital mixing, some of these $\sim 7.7\%$ of particles will temporary join the constant fraction of $\sim 1.1\%$ to build the measured $\sim 3 \%$ compressive fraction. To estimate the duration of their compressive episodes, we compute the velocity associated to the angular momentum at the boundary of the compressive zone and derive a mean orbital period of $53 \Myr$. This means that the particles enter and leave the central compressive region on a timescale of $\sim 50 \Myr$. (Note that the free-fall time of this region is $t_\mathrm{ff} = 26.3 \Myr$ and that $2 t_\mathrm{ff}$, i.e. the time needed to get to the center and exit the region, is approximately the value we derived.) Therefore, one should not expect a high fraction (of the non-constantly compressive particles) to present a \emph{lus} greater than $50 \Myr$.

\begin{figure}
\plotone{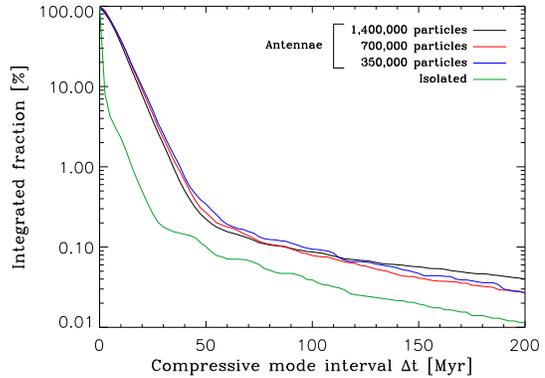}
\caption{Distributions of the longest uninterrupted sequence in compressive mode. Integration has been performed for $t \in [-100, 1000] \Myr$ for the Antennae and for the same time ($1100 \Myr$) for the isolated case. Again, lower resolutions presents the same results ($\sigma = 0.48$ and $\sigma = 0.43$ for 700,000 and 350,000 particles respectively).}
\label{fig:period_resolution}
\end{figure}

Fig.~\ref{fig:period_resolution} reflects the points described above (for \emph{lus} only). It plots the integrated mass fraction $M(>\Delta t)$ which experienced a continuous compressive event longer than $\Delta t$, for the Antennae case as well as for an isolated progenitor. These distributions are normalized by considering the non-constantly compressive particles only, which means that $M(\Delta t = 0) = 100\%$. As expected, in the isolated case, $\sim 8\%$ of the mass experience a compressive episode longer than $2.5 \Myr$ while only $0.1 \%$ will be in this mode for more than $50 \Myr$.

For the Antennae case, the distribution is well fitted by exponential laws $M(>\Delta t) \propto \exp{(\Delta t / \tau)}$ in the time intervals $\Delta T_1 = ]0,50] \Myr$ and $\Delta T_2 = [50,100] \Myr$. For the \emph{lus}, we obtain $\tau_1 \simeq 8.0 \Myr$ and $\tau_2 \simeq 50.2 \Myr$ for the $\Delta T_1$ and $\Delta T_2$, respectively. The isolated case presents similar slopes.

\begin{figure}
\plotone{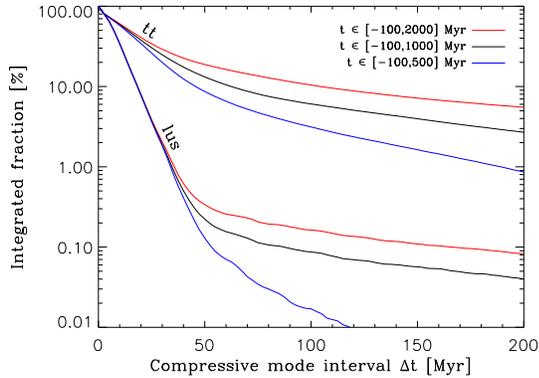}
\caption{Distribution of the longest uninterrupted sequence (\emph{lus}) and total time (\emph{tt}) spent in compressive regime, for various integration times. If a strong dependence is visible for long episodes, the short events ($< 50 \Myr$) remain well-defined, whatever the integration time is. We note that the position of the knee is not affected.}
\label{fig:period_integration}
\end{figure}

One has to be careful before jumping to conclusions. These distributions depend on the time of integration, i.e. the time of calculation, during which one allows the particles to enter into the compressive regime. For an integration over a long time, the distributions would be flatter, because a higher number of particles could experience one or more compressive episodes. To check the validity of our results, we plot these distributions for various integration times (from $600 \Myr$ up to $2.1\Gyr$) in Fig.~\ref{fig:period_integration}, for the Antennae case. In addition to the \emph{lus}, the total time (\emph{tt}) spent in compressive mode (i.e. the sum of all the compressive events) is also shown. The position of the knee (at $\sim 50 \Myr$) does not vary but, as expected, we note a large difference in the \emph{tt} distribution (up to a factor 6). As the purpose of this work is to concentrate on the first $100 \Myr$ of these modes, this variation does not strongly affect the conclusions presented here. For most of the particles, the \emph{lus} begins at the first pericenter passage and thus, the associated distribution remains stable to a change of integration time, provided it includes the two passages (peaks in Fig.~\ref{fig:histo_resolution}). It is also important to note that a \emph{tt} distribution integrated over a long period is physically less relevant as it is punctuated by longer and longer extensive episodes, which could blur the effects of the previous compressive events.

\begin{figure}
\plotone{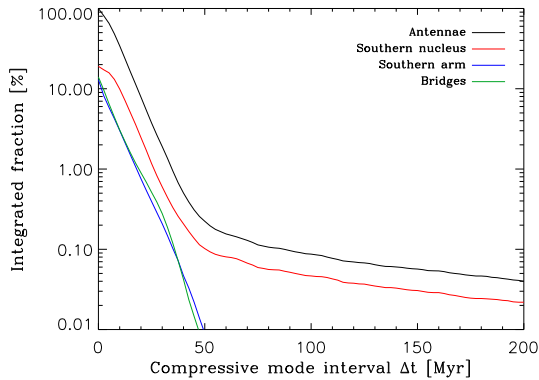}
\caption{Distribution of time intervals spent in compressive mode (\emph{lus} only) for different regions of the Antennae galaxies. The transient features (tails and bridges) present very short periods (always less than $50 \Myr$), while the nuclei have a long period regime, too. Thanks to the symmetry of our Antennae model, the structures of the southern galaxy only are presented.}
\label{fig:period_regions}
\end{figure}

For the Antennae, one could explain the short periods with the existence of transient features and link the longer episodes to the intrinsic compressive areas, like in the isolated case. Fig.~\ref{fig:period_regions} confirms this assumption by showing the distribution of the compressive periods for various regions of the merger. It distinguishes three main areas: the nuclei, the tidal tails and the bridges. As expected, the transient structures like the tails and the bridges show very short compressive periods while the central parts (nuclei) present the knee at $\sim 50 \Myr$, between short and longer periods. Therefore, the intrinsic compressive regions are the only ones to experience compressive episodes of few $100 \Myr$. A shorter period is also visible and matches the associated distributions for the isolated progenitors. Hence, one can tell apart the effect the orbital mixing (short and long events) and the role of the merger (short periods only).

\subsection{Compressive history}
\label{sec:history}

\begin{figure}
\plotone{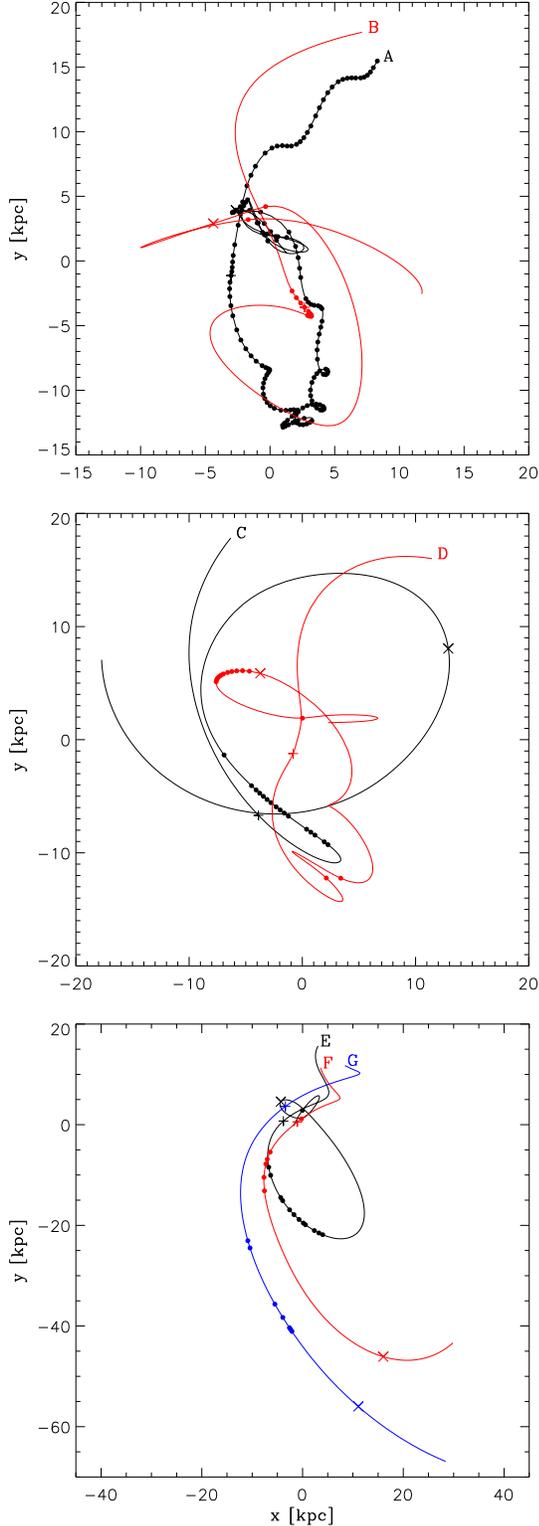}
\caption{Orbits of the characteristic particles (in the orbital plane) for the inner regions (top), intermediate zone (center) and large scale (bottom). Dots show the compressive events. The two pericenters are marked with crosses ($+$ and $\times$ respectively) on every orbit. Particles A and B stay in the vicinity of the central region. C goes into one of the bridges, D falls back in the merger remnants in the merger phase ($t=300\Myr$). Particle E is ejected to the southern tail by the first encounter but falls back into the nucleus before the second. F stays in the tail and G is part of the TDGs region. One position (i.e. one snapshot) corresponds to $2.5 \Myr$.}
\label{fig:orbits_particles}
\end{figure}

The typical period of $50 \Myr$ reported above matches the time scale of star cluster formation. In addition, the positions of the compressive regions matches amazingly the location of observed young structures. These clues underline the role of the tides in the birth of clusters or TDGs. However, one may ask how the statistics presented above reflect the actual tidal history of a single object. Indeed, the \emph{lus} concept (introduced for statistical purposes) only tells us about one compressive event, while a star cluster orbiting in the Antennae is expected to experience many of them. Therefore, we propose a new description of the local tidal field, by selecting a few particles in each region of interest of our Antennae model: the central part, the bridges, the tail in the vicinity of the nuclei and the TDGs region. The orbits of these particles are plotted in Fig.~\ref{fig:orbits_particles}. 

This new approach allows to retrieve the tidal history of individual orbits and, with the statistics obtained before, to get a broader picture of how compressive modes evolve. In other words, Fig.~\ref{fig:orbits_particles} shows some realizations of the statistics we presented. These examples define realistic boundary conditions that will be used for the simulation of star clusters. Note that each particle represents a special case among the mean behavior of the associated region. Therefore, one can pile up snapshots during which a body stands in compressive mode, and consider that as a compressive event, even if it is punctuated by \emph{short} extensive episodes. That is, most of the compressive occurrences follow the global history of the merger, as related on Fig.~\ref{fig:morphology} and \ref{fig:histo_resolution}.

Indeed, the particles spotted in the bridges (B, C) or the tails (F, G) show their major compressive event after the first pericenter. One can also track the fall-back of the particles into the nucleus (E) or simpler, the long period of a particle that stays in the center for the entire simulation (A).

This figure helps to distinguish the major families of compressive histories one could find in a merger. The behavior of these selected particles matches the statistics derived above, in terms of position in space and time, and time scale.

\section{Parameter survey}

\begin{figure*}
\plotone{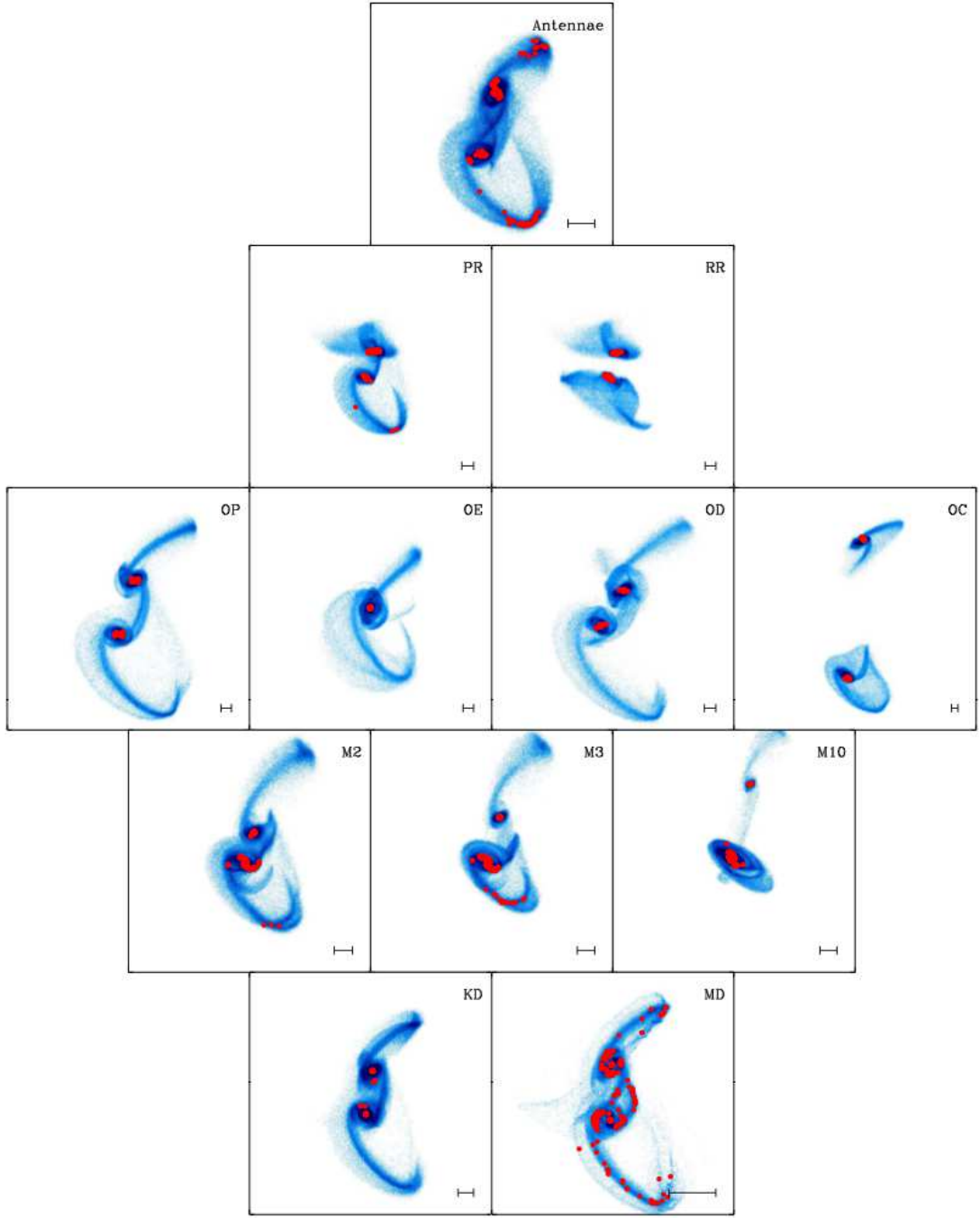}
\caption{Morphology of the 12 models at (approximately) the same dynamical stage, between the first two passages. The same line of sight has been used to ease the comparisons. The images are centered on the center of mass of the entire system (dark matter included). On each image, the black line corresponds to $10 \kpc$. Red dots denote the particles standing in fully compressive mode.}
\label{fig:morphology_survey}
\end{figure*}

We performed a series of simulations, each with a physical parameter set to a different value compared to the reference Antennae run, with a view to quantify the impact of these parameters on the statistics of the tidal field. This section describes the various models; Table~\ref{tab:survey} sums up the main results, and Fig.~\ref{fig:morphology_survey} shows the morphologies and the spatial distribution of the compressive regions after the first pericenter passage.

\begin{deluxetable}{llrrr}
\tablecaption{Parameter survey\label{tab:survey}}
\tablehead{\colhead{Model name} & \colhead{Key parameter} & \colhead{Peak 1\tablenotemark{a}} & \colhead{Peak 2\tablenotemark{b}} & \colhead{$\tau_1$\tablenotemark{c} [Myr]}} 
\startdata
\multicolumn{2}{l}{Antennae (see \S \ref{sec:antennae})} &	11.5\% &	11.5\% &	8.0\\
\hline
\multicolumn{5}{l}{Prograde-retrograde encounter (see \S \ref{sec:spin})}\\
PR &		1 retrograde disk &	7.5\% &	14.0\% &	8.0\\
RR &		2 retrograde disks &	7.0\% &	18.5\% &	8.1\\
\hline
\multicolumn{5}{l}{Orbits (see \S \ref{sec:orbit})}\\
OP &		high eccentricity &	10.0\% &	9.5\% &	7.7\\
OE &		low eccentricity &	14.5\% &	8.5\% &	8.3\\
OD &		distant pericenter &	3.5\% &	9.5\% &	6.0\\
OC &		close pericenter & 	15.5\% &	15.5\% &	7.0\\
\hline
\multicolumn{5}{l}{Mass ratio (see \S \ref{sec:mass})}\\
M2 &		2:1 mass ratio &	8.5\% &	5.5\% &	7.2\\
M3 &		3:1 mass ratio &	5.5\% &	6.5\% &	6.9\\
M10 &		10:1 mass ratio &	1.5\% &	5.5\% &	6.7\\
\hline
\multicolumn{5}{l}{Progenitors model (see \S \ref{sec:halo})}\\
KD &		other model &		4.5\% &	1.5\% &	2.5\\
MD &		other model &		10.0\% &9.0\% &	3.2
\enddata
\tablenotetext{a}{Mass fraction in compressive mode added at the first pericenter passage (compared to the isolation stage).}
\tablenotetext{b}{Mass fraction in compressive mode added at the merger phase.}
\tablenotetext{c}{Exponential timescale of the \emph{lus} distribution for the time interval $\Delta T_1$ (see Section~\ref{sec:charcteristic_times}).}
\end{deluxetable}

\subsection{Prograde - retrograde encounter}
\label{sec:spin}
It has been clear since \citet{Toomre1972} that the large-scale morphology is linked to the initial parameters of the stellar disks. Indeed they showed that, e.g., the intrinsic angular momentum of a galaxy (spin) could be the key factor in the creation of the tidal tails. We propose a similar exploration of the influence of the spin, for disks embedded in a dark matter halo.

The reference Antennae calculation is a prograde-prograde (PP) encounter, i.e. both disks' spins are coupled with the orbital angular momentum of the galaxies. This configuration allows the growth of the two observed tails. In this model, $\sim 30\%$ of a single disk mass is ejected to form the associated tidal tail. (This mass loss was noted in earlier studies, see e.g. \citealt{Barnes1988}.) Varying the spin of one or both of the disks allows to change the tidal structures created during the encounter. For a retrograde galaxy (spin anti-aligned with the orbital angular momentum), the bound mass remains approximately the same. This affects the ensuing morphology, but also the dynamical mass. Indeed, in a retrograde-retrograde configuration (RR), the motion of the center of mass of each galaxy shows a larger separation ($\simeq 4 \kpc$) than for PP, and thus an orbital period $\sim 100 \Myr$ longer.

We first set a prograde-retrograde (PR) simulation by flipping one of the two disks. Only the prograde galaxy creates an arm while the mass of the other disk remains almost constant during the encounter (see the second row of Fig.~\ref{fig:morphology_survey}). At the second passage (i.e. just before merger), the amount of material in that disk which can experience a compressive mode is much higher.

A merger event is violent and affects entirely what is left of the disk, \emph{after the first passage}. In a first approximation, one can consider that the reaction of the disk material would be the same, whatever orientation the spin takes. However, for prograde galaxies, a significant fraction of this mass has been ejected in the form of tidal tails. If a given fraction $X$ of the bound mass undergoes a compressive mode, the equivalent fraction for the total mass (disk + tails) must be lower. In other words, a retrograde disk provides more mass to enter in a compressive mode, thus the equivalent fraction is higher than for a prograde configuration.

For the Antennae run (PP), $14\%$ of the stellar mass is in compressive mode at the second pericenter passage ($t = 300 \Myr$). Each disk has lost $30 \%$ of its initial mass into tails. Therefore, $X = 0.14 / (1.0 - 0.3) = 20 \%$ of the bound mass stands in compressive mode. For a retrograde galaxy, we expect $\sim 20\%$ of the mass to experience compressive tides. In the PR case, we have one prograde disk ($14 \%$) and one retrograde progenitor ($20 \%$) of equal mass. Therefore we expect $(20\% + 14\%) / 2 = 17 \%$ of the mass in compressive mode. Because of the change in dynamical mass, the second pericenter passage of PR model should occur $\sim 50 \Myr$ after those of the PP configuration.

Using the same arguments, we can predict that the RR case would have up to $20\%$ of its mass in compressive mode (because no tail has been created), and this, $\sim 100 \Myr$ after the PP case.

\begin{figure}
\plotone{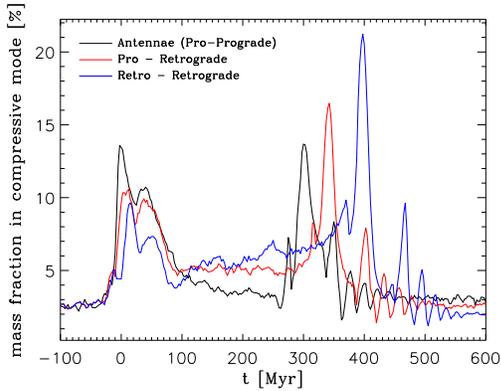}
\caption{Evolution of the mass fraction in compressive mode. Relative amplitude of the peaks and time shifting can easily be explained by mass loss consideration. See text for more details. Note that the retrograde-prograde configuration (not shown here) almost matches the prograde-retrograde one, as our progenitors are similar.}
\label{fig:histo_ppprrr}
\end{figure}

Fig.~\ref{fig:histo_ppprrr} shows the results. The initial $\sim 3\%$ offset and the first peak correspond to non-altered disks and thus, are common to the three configurations. When the two galaxies move apart ($t \in [0,250] \Myr$) the decreasing rate of the mass in compressive mode depends on the geometry and remaining mass in the central regions. The second passage, as expected, spreads from $t = 300 \Myr$ for the Antennae case to $t = 400 \Myr$ for the RR model. The maximum compressive fraction is in good agreement with the argument discussed above. The last small peaks stem from the oscillations of the distance between the two nuclei which finally merge into a single ellipsoid. The three configurations show these spikes, as the final merger phase does not depend on the external tails. Once the overall potential well stabilizes, the compressive fraction remains constant, within fluctuations of similar amplitude than the initial noise.

Even if the compressive peaks are shifted in time and have different amplitudes, the overall behavior of these curves remains the same, whatever the orientation of the spins is. However, it is important to remark that in all these cases, the impact angle between the two progenitors is the same (modulo $180^{\circ}$). We also run simulations with other angles to check this. Even a rotation of $90^{\circ}$ of one of the progenitors does not affect the mass fraction in compressive mode. Only the time interval leading up to the second peak increases by $50 \Myr$.

\begin{figure}
\plotone{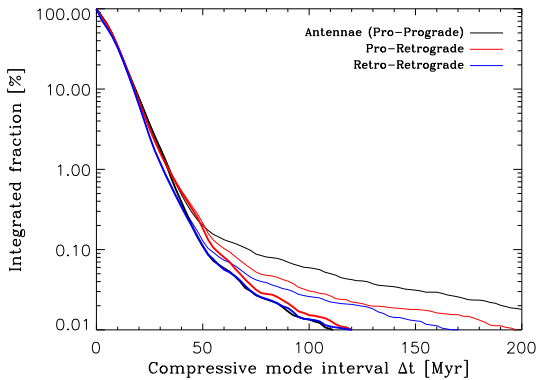}
\caption{Distribution of the \emph{lus} in compressive mode for the PP (Antennae), PR and RR cases. For the thin curves, the integration time has been set from $100 \Myr$ before the first passage, to three times the pseudo-period (i.e. the time-lapse between the main two peaks). In all cases, the short periods and the knee at $50 \Myr$ are similar. For the thick curves, integration has been stopped just after the merger event (at $t = 375, 440$ and $500 \Myr$ for the PP, PR and RR cases respectively).}
\label{fig:period_ppprrr}
\end{figure}

A visual inspection of the width of the peaks in Fig.~\ref{fig:histo_ppprrr} allows to predict that the short time scale in compressive mode (\emph{lus} distribution) would not change with the spin of the galaxies. Indeed, Fig.~\ref{fig:period_ppprrr} shows that the slope derived by exponential fits in section~\ref{sec:charcteristic_times} remains the same, as well as the position of the knee at $\sim 50 \Myr$. 

Analyzing the longer periods is more difficult, as they strongly depend on the integration time. Because of the mass loss of the prograde galaxies, the dynamics of the three systems are not the same. We define a pseudo-period as the time-lapse between two major peaks of compressive mass fraction (see Fig.~\ref{fig:histo_ppprrr}). For the thin curves of Fig.~\ref{fig:period_ppprrr}, we set arbitrarily the integration period to three pseuso-periods, i.e. $1000, 1150$ and $1300 \Myr$ for the PP, PR and RR cases. The difference in the final mass fractions visible in Fig.~\ref{fig:histo_ppprrr} is integrated for a long time. The RR configuration, which shows only $\sim 2\%$ of the mass in compressive mode \emph{after} the merger phase, has a lower \emph{lus} distribution than the PP one (which equivalent value is $\sim 3\%$).

However, in order to get a fair comparison between the models, one can link the integration time to a dynamical stage, and not a period based on the orbit. By stopping the integration just after the merger (Fig.~\ref{fig:period_ppprrr}, thick curves), we deal with three objects showing the same dynamical stage in the Toomre sequence \citep{Toomre1972}. In that case, the comparisons regarding time are much more relevant. Indeed, the \emph{lus} distributions are very similar, even for long periods, showing that the spin of the progenitors does not influence the timescales of the compressive tides (see Table~\ref{tab:survey}).

\subsection{Orbits}
\label{sec:orbit}
The next part of this survey focuses on the influence of the progenitor orbits, keeping all the other parameters (e.g. the spin or the mass ratio) as in the Antennae reference. Because the galaxies are not point-masses, it is not possible to use a Keplerian description of their orbit. However, before the first pericenter passage, the motion of the centers of mass of the disks can be approximated as a two-body problem.

To ease the comparisons, in the following we will refer to the equivalent Keplerian orbits. Table~\ref{tab:orbits} describes the parameters (eccentricity $e$ and pericenter distance $d$) of a set of representative orbits. Because of the orbital decay due to dynamical friction, this two-body approximation is no longer valid after the first passage.

\begin{deluxetable}{lrrl}
\tablecaption{Orbital survey\label{tab:orbits}}
\tablehead{\colhead{Model} & \colhead{$e$} & \colhead{$d$ [kpc]} & \colhead{Remarks}} 
\startdata
Antennae & 0.96 & $5.8$ & reference model \\
OP\tablenotemark{*} & 1.10 & $5.8$ & almost parabolic \\
OE & 0.80 & $5.8$ & low eccentricity \\
OD & 0.96 & $13.2$ & distant encounter \\
OC\tablenotemark{**} & 1.50 & $3.5$ & close encounter
\enddata
\tablenotetext{*}{Orbital decay makes this orbit almost bound.}
\tablenotetext{**}{High eccentricity is required to avoid a central collision (due to orbital decay) that would hide the effect of the initially finite pericenter distance.}
\end{deluxetable}

\begin{figure}
\plotone{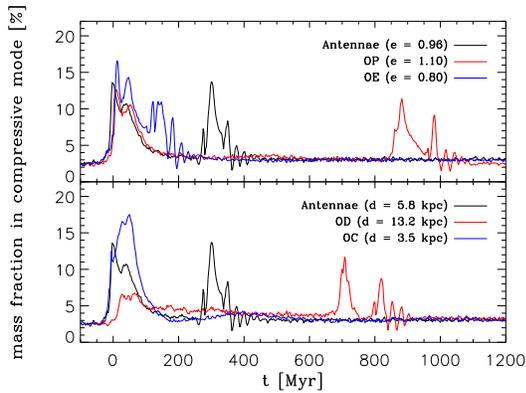}
\caption{Effect of the eccentricity (top) and the pericenter distance (bottom) on the mass fraction in compressive mode. A low eccentricity concentrates the compressive events during the only passage, while a higher value involves a third passage (at $\sim 975 \Myr$). A close encounter allows a high mass in compressive region.}
\label{fig:histo_orbits}
\end{figure}

The evolution of the mass fraction in compressive mode is given in Fig.~\ref{fig:histo_orbits}, with the Antennae for comparison. 

At the first passage, the OP run shows a similar behavior to the reference. However, its large eccentricity gives it a long orbital period, which leads to a second passage\footnote{The eccentricity of 1.10 is not large enough to overcome dynamical friction and leave the galaxies unbound.} delayed by $\sim 600 \Myr$. A third encounter occurs just before the merger phase and its oscillatory mode. \emph{A contrario}, a small orbit like OE brings back the galaxies very quickly and thus induces the merger while the transient tidal features still exist.

For a distant passage (OD), the overlap of the potentials which leads to a compressive zone is much smaller and thus, the corresponding mass fraction is lower ($\sim 6\%$ instead of $14\%$ for the Antennae). The orbital decay brings the two progenitors back at $t \simeq 700 \Myr$, and the major compressive event occurs. In the close passage case (OC), almost the entire mass is available to enter into the compressive regime. Therefore, the first pericenter peak is much higher ($\sim 18\%$). Dynamical friction plays a crucial role as it prevents the progenitors to fly away from each other. As a consequence, only one long compressive episode is visible.

Here again, we set the end of the integration time to the merger stage, to have the same dynamical phase for all the orbits. As expected, the short period ($\tau_1$) does not depend on the eccentricity. However, we note a slight influence of the pericenter distance (see Table~\ref{tab:survey}). The distant encounter presents a shorter period ($\sim 6 \Myr$ instead of $8 \Myr$) because the passages are much quicker, which is visible with the width of the peaks in Fig.~\ref{fig:histo_orbits}. For the OC case, the merger stage is too close to the pericenter. Therefore, the characteristic timescales derived are not physically relevant for a comparison.

The broad conclusion of this survey is that a change of the orbital eccentricity has a mild effect on the mass loss and the orbital period. Only the pericenter distance influences significantly the importance of the compressive modes. To be more specific on this last point, we explore in the next subsection the trend of compressive modes with impact parameter.

\subsection{Fly-bys versus mergers}
\label{sec:intermerger}
Star clusters also form in galaxies that are undergoing milder interaction, as starburst galaxies show different levels of intensity. To see how the statistics of compressive modes is affected by the character of the interaction, we varied the impact parameter $d$ of the two progenitors, ranging from one to 14 disk scalelengths $R_d$, keeping all other quantities unchanged. In this fashion, the simulations cover the full range from merging (bound orbits) to mildly interacting galaxies (hyperbolic orbits, or fly-bys). Nevertheless, our parameter survey remains limited in scope, as we do not include high-velocity encounters with low impact parameters, which would also produce fly-bys. We chose to keep the initial relative velocity ($\sim 200 \U{km.s^{-1}}$) constant, in part because this value seems of the right order for field galaxies (as opposed to $\sim 1000 \U{km.s^{-1}}$ for cluster galaxies); and also because higher velocities would lead to shorter timescales, when effects of tidal fields may be appropriately treated in the limit of a shock \citep[see][]{Spitzer1987, Gieles2007}. However, let us note for the sake of clarity that a fast (shock) encounter will trigger a strong response from the gas whenever the impact parameter is so small that the disks overlap.

\begin{figure}
\plotone{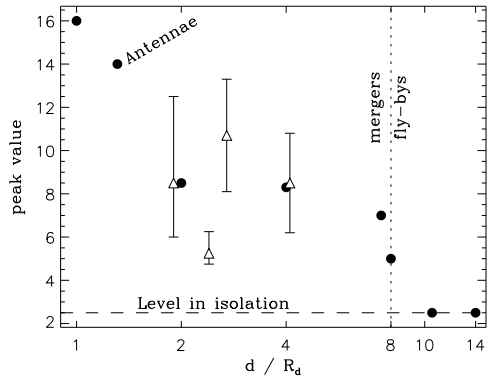}
\caption{Mass fraction in compressive mode at the first pericenter passage versus the impact parameter $d$ normalized to the disk scalelength $R_d$. Filled circles correspond to compressive modes for our runs, while open triangles are SFR derived from the hydrodynamical simulations of \citet[Fig.~21]{diMatteo2007}.}
\label{fig:intervsmergers}
\end{figure}

We graph on Fig.~\ref{fig:intervsmergers} the mass fraction in compressive mode at the first pericenter passage as a function of $d / R_d$. Each filled circle on the figure represents one simulation. The vertical dotted line marks the minimum distance for the turnover from merger to fly-by interactions. Crucially for non-merging galaxies, the mass fraction remains at the same level as that of the progenitors. When $d / R_d$ is decreased, the fraction increases slowly but peaks sharply when $d / R_d \sim 1$. As a result, if we were to match compressive modes with star formation episodes, we would predict enhanced formation of stellar associations with decreasing impact parameter, and no enhancement for fly-bys. It is interesting to contrast this with simulations including gas. Fig.~21 of \citet{diMatteo2007} shows an enhanced formation when compared to fly-bys for all impact parameters in the range $d / R_d \sim 2$ to 6 ($R_d$ is taken equal to the parameter $a_{\star}$ of their Miyamoto-Nagai models). However, these authors also found a trend whereby the star formation rate (SFR) \emph{increases} slightly with impact parameter, which may be interpreted as a result of more gas falling in the deep potential of the final merger phase. If we focus on their data for late-type (disk) galaxies, this trend is not incompatible with a constant maximum SFR, irrespective of whether the orbits are retrograde or prograde. This is very akin to our own data for mergers in the same parameter range. On Fig.~\ref{fig:intervsmergers}, we display as open symbols their data for spiral-spiral mergers (denoted L-L by them). The good overlap of values suggests compressive modes may be very important during a galaxy merger. On the contrary, our data for fly-bys indicate no enhancement whereas hydrodynamical runs show an enhancement of a factor two or more. This is indicative that a fine treatment of the hydrodynamics is essential to predict a global SFR, while gravity alone would not be enough. Note importantly that we limited our study to scales of $40 \U{pc}$ or larger so that this conclusion only applies to relatively massive stellar associations.

\subsection{Mass ratio}
\label{sec:mass}
One key parameter in the study of interacting pairs of galaxies is the mass ratio \citep{Naab2003, Bournaud2005, Naab2006, Johansson2009}. Most of the observational and theoretical facts set apart major mergers (mass ratio greater than 3:1) and minor mergers. To see how this could influence the compressive tidal mode, we introduce three new models, in addition to the Antennae (of mass ratio 1:1) reference: another major merger (2:1), a minor merger (10:1) and an intermediate mass ratio (3:1).

In this study, the mass ratio refers to the total mass (stars + dark matter) and thus, is closer to the cosmological definition than to the observational one \citep[see][]{Stewart2009}. The galaxies have been scaled to keep the density and the virial ratio constant. This choice permits to maintain the intrinsic compressive region at $\sim 3\%$ of the disk mass.

On the one hand, the small intruders are strongly affected by their passage in the vicinity of the main disk. However the tidal effects they undergo are mainly disruptive (extensive) and thus, only their central region remains compressive. On the other hand, the main disk stays coherent and its mass loss is much less important than for its companion. It is important to note that the passage of the small galaxy through the large disk induces radial perturbations which rapidly gives birth to spiral arms and then, a bar. These transient features generally allows the formation of local cores in the gravitational potential, and thus, the raise of local compressive fields.

As a consequence, the mass fraction in compressive mode is higher for the main disk, when a small intruder hits it, than in isolation. However, the intruder itself does not show a strong change in its compressive behavior. Finally, the total mass fraction in compressive mode should be low when small galaxies are involved.

\begin{figure}
\plotone{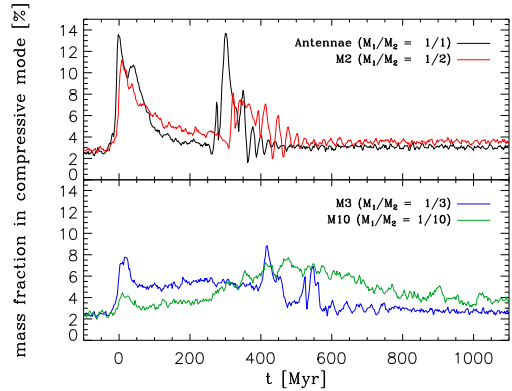}
\caption{Evolution of the mass fraction in compressive mode for major (top panel) and minor (bottom panel) mergers. The low mass progenitors present much less mass fraction in compressive mode than the higher mass galaxies.}
\label{fig:histo_mratio}
\end{figure}

Indeed, our results (see Fig.~\ref{fig:histo_mratio}) show that this fraction is considerably lower for a small mass ratio (3:1 and 10:1) than for the major mergers (1:1 and 2:1). We detected features common to all the simulations, such as the presence of a peak to denote the creation of the tidal tails, or an oscillatory phase for the later steps of the merger. As the different masses change the orbital period, the main peaks are delayed for the low mass progenitors.

In the cases where spirals and bars are created, the compressive mass fraction presents a plateau during the lifetime of these structures. The associated timescales are similar to those of the tidal tails and bridges (see Fig.~\ref{fig:period_regions}). As a result, the exponential timescale $\tau_1$ (of the \emph{lus} integrated until the same dynamical stage) is slightly shorter for the minor mergers than for the major ones (see Table~\ref{tab:survey}).

\subsection{Further compound models}
\label{sec:halo}
The last step of our survey deals with different compound galaxy models. Here, the main goal is to quantify the influence of the halo on the evolution of the merger, both in terms of the total mass of the halo, i.e. its truncation radius, and its mass profile. This last point should bear on the growth of bridges and the orbits of the nuclei, since a more peaked dark matter halo will concentrate more mass near the center. It is therefore of interest to compare the statistics of compressive modes with models where the halo does not have an isothermal core, as does the reference Antennae model discussed extensively in Section~\ref{sec:antennae}.

At first sight, it should have sufficed to change the halo mass profile and keep the stellar content as in the reference model to ease comparisons. However, the stability of the disk especially requires the coherent adjustment of all three components. We made use of equilibrium distribution functions from the literature to create new composite models, all scaled to have a circular velocity of $220 \U{km.s^{-1}}$ at the solar radius.

The first model is a Kuijken-Dubinski model (hereafter KD) of a Milky Way-like galaxy \citep[][model MW-A]{Kuijken1995}. It uses a lowered Evans distribution function for the halo ($\approx 1.3$ times more massive than the Antennae's), a King model for the bulge and a 3D version of a Shu disk, truncated at the same radius than the reference model, but $\approx 1.5$ times more massive. As mentioned before, the mass fraction in compressive mode in equilibrium should vary from one model to the next, as this fraction is linked to the form of the potential. The total density in the disk built by the overlap of the disk, bulge and halo sets the local Jeans length over which a fragmentation mode may grow and form cores of compressive tides. Because the Jeans length will increase as the density decreases, a more extended mass profile, resulting in a shallower potential, is expected to yield a lower mass fraction in compressive modes. As the KD model is more extended than the reference Antennae model, we derived a mass fraction of $\approx 0.8 \%$ instead of $\sim 3\%$.

The KD model goes through the by-now usual evolutionary steps: a first passage, followed by the formation of the bridges and tails, before merging completely. Because the mass ratio between the two progenitors is the same as the reference model, the Keplerian orbits of both models match well. Despite the fact that the KD model is more extended, differences in the orbits attributable to dynamical friction are not significant. Indeed, the pseudo-period of $\sim 300 \Myr$ (as defined in Section~\ref{sec:spin}) is almost the same. However, the low average density of the KD model leads to a more effective dispersion of the disks material, which results in depopulated nuclei. Bearing in mind the considerations of the previous paragraph, one may deduce that the compressive mass fraction will rise, at the first passage, by a factor similar to the Antennae model, but by a smaller amount at the later stage of merger, because the core region has been more effectively depleted (see Table~\ref{tab:survey}).

Even if KD and the reference model share a similar rotation curve, the energy and angular momentum of individual orbits are not the same. We computed shorter timescales ($\simeq 2.5 \Myr$) for the compressive tides, showing a more efficient orbital mixing in isolation as well as during the merger. A detailed explanation for this seems out of reach as too many parameters have been changed. However, the exponential distribution of the compressive mode still has a clear signature, but with shorter e-folding times.

The next model is taken from \citet[][hereafter MD]{McMillan2007}. The halo component is a massive NFW profile with a scale length of $\approx 12 \kpc$. The bulge is a \citet{Hernquist1990} profile, while the disk decays exponentially with cylindrical radius over a characteristic scale $\approx 2.8 \kpc$. Note that the dark matter halo is much more massive and extended than in the previous models to the extent that the halos would overlap if set on the same orbits used for the Antennae model. Indeed, the entire disk sits within on scale length of the NFW profile (where $\rho \propto r^{-1}$). Even if this central region is intrinsically tidally extensive, the addition of other potentials (disk and bulge) could more easily build compressive areas than in the external zones ($\rho \propto r^{-3}$). As before, one should not expect the intrinsic mass fraction in compressive mode to be the same as for our reference. In this case, the fraction is $\approx 0.6\%$. We noted that not only the central part showed compressive tides, but so did a number of small areas distributed all over the disk. These small zones are clearly compressive because of a transient distribution of the matter at small scales, and thus do not have long compressive episodes. The same occurs in the course of the merger.

Here again, the time-evolution of the mass fraction presents the usual sequence of peaks. We note that the high halo mass (24 times that of the disk) reduces considerably the orbital period. In addition, it increases the role of the dynamical friction, as both galaxies stand inside each other's halo. As a consequence, the merger stage occurs earlier than for the other models (i.e. $\sim 200 \Myr$ only after the first passage). In the inner part of the NFW potential, the overlap of the two disks induces many compressive areas. In Fig.~\ref{fig:morphology_survey} (last panel), one observes much more compressive particles in the tidal structures (bridges and tails). Indeed, between the pericenter passages, the compressive regions in the arms do not have enough time to dissolve. Consequently, the mass experiencing compressive modes remains fairly high in the MD run, while it has time to decrease in the other Antennae - and KD simulations. This also imprints the timescales. Indeed, the combination of both the transient small compressive areas and the short orbital period limit the exponential timescale of compressive mode to $\simeq 3.2 \Myr$. The results listed in Table~\ref{tab:survey} reflects these points.

In conclusion, even if a modification in the mass distribution of the galaxy model does induce changes in the statistics of compressive modes, their general behavior (e.g. location in the tails and bridges, exponential timescale, timing) remains overall unchanged, and, especially, the statistics of compressive modes at the time of the final merger is relatively robust against details of the models.

\section{Discussion}

\subsection{Statistics of the tidal field}
The gravitationnal tidal field of galaxies is often presumed to disrupt smaller bodies orbiting in their vicinity. This point of view stems from a wealth of observational data of on-going disruption of tidal dwarf galaxies, as well as strong theoretical underpinning going all the way back to Roche and the notion of an outward stretch operating on finite-size bodies. But a close inspection of the full tidal field tensor leads one to conclude that in several realistic situations, such as in the core of flat-top density profiles or the mid-point of spiral disks, the tide will work in reverse and act to shelter structures from dissolution. Because the character of the tidal field is sensitive to the presence of clumps or other sub-structures, it seems natural to seek out quantitatively the evolution of the tides in the context of a major galaxy mergers, when large structures and smallish clumps will form in great numbers. A key question that we have addressed in this paper is whether compressive tidal modes could widely influence the early life of star clusters or TDGs, both in terms of their location in the merger, and their duration in time.

The reference Antennae model discussed earlier shows that fully compressive tidal modes spread throughout the merger, including in key areas such as the nuclei and the tidal tails. The statistics of the duration of these modes reveal that most of them last long enough to impact on the formation of bound substructures on a scale of $\sim 200 \U{pc}$ or smaller. This conclusion was confirmed in a parameter survey of galaxy mergers, which showed that the shape of the tidal mode distribution function is by far unaffected by details of the systems undergoing a merger (spin-orbit coupling, mass ratio, impact parameter, and so on).

Broadly speaking, compressive tidal modes always develop to reach the same level of intensity in all the simulations that we have performed. However, some trends have been highlighted. First of all, the mass fraction entering compressive modes scales with the available mass: a low-mass progenitor undergoing a collision yields a low mass-fraction of compressive tidal modes. This is clear from the spin- and mass ratio explorations. Secondly, an encounter with a small pericenter distance brings more material to the collision in a smaller volume. As a result, the number of mass elements experiencing compressive tidal modes is significantly higher for close passages.

\subsection{Tidal field and star formation rate}
For every interaction leading to mass exchange and possibly a merger, the fraction of bodies experiencing a compressive tidal field rises by a factor of at least 4, and up to 13 in the extreme case of a face-on collision, when compared with galaxies in isolation. These figures match well with the results of \citet{diMatteo2008} who presented a study of the SFR over a large set of mergers, using both SPH and sticky particles techniques. We also remark that the starburst activity derived from their simulations would not last longer than $\sim 500 \Myr$, which matches well with our results. This provides indirect evidence of a close link between fluctuations in the global galactic gravitational field, and the small-scale volumes where stars are presumed to form \citep[recall Fig.~2 of][for another, complementary argument]{Renaud2008}.

\Citet{diMatteo2008} highlighted that retrograde encounters enhance the SFR of mergers, which is again in agreement with the evolution of the tidal field for this case (cf. models PR and RR of Section~\ref{sec:spin}). However, they noted a negative correlation between the pericenter distance and the starburst, in the sense that close passages correspond to \emph{low} SFR, for mergers \emph{only}. It is yet unclear to what extent this trend can be attributed to the response of the gas in the early stages of the merger. Because the statistics reported here do not account in details for the sequence of events that take place, we can not be specific about the likelihood of forming stars at a precise time. A clearer picture will emerge once high-resolution hydrodynamics is included in the dynamical scenarios that we have reported here.

\subsection{Age distribution function}
In all merger simulations that we have carried out, a double exponential law describes well the compressive mode distribution in age, for both the longest uninterrupted sequence (\emph{lus}) and the total time (\emph{tt}). We note that the longest periods of the \emph{lus} statistics correspond to orbits near the centre of the galaxies (in isolation as well as during the interaction) while the transient features that give rise to the shorter periods of the \emph{lus} distribution were found in the merger-induced tidal structures (such as bridges and tails).

In Section~\ref{sec:history}, we showed that the history of the gravitational tidal field along individual orbits is very complex. Indeed, to compute the net effect of the tides on a star cluster over time via a statistical approach, one should combine \emph{both} the \emph{lus} and \emph{tt} distributions. This is so because the tidal field changes character (compressive or extensive) repeatedly along most orbits (see Fig.~\ref{fig:orbits_particles}).

\begin{figure}
\plotone{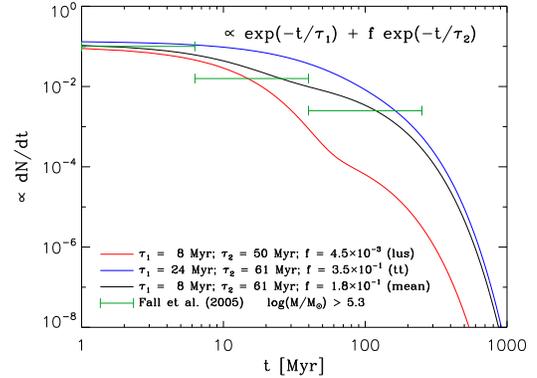}
\caption{The \emph{lus} and \emph{tt} distributions (red and blue lines, respectively) of compressive modes for the Antennae model. The distributions consist in double exponential laws fits to the short and long timescales ($\Delta T_1$ and $\Delta T_2$). The black line represents a combination of both cases, using the short timescale from the \emph{lus} and the longer one from \emph{tt} (see text for details). The green points with horizontal binning are HST data lifted from \citet{Fall2005}.}
\label{fig:age}
\end{figure}

To illustrate this further, consider once more the case of the Antennae model. Roughly speaking, the statistics of compressive modes for that case can be split in two phases: a first phase in the time interval $\Delta T_1 = ]0,50] \Myr$, during which the compressive modes are mostly uninterrupted; and a second phase $\Delta T_2 = [50,100] \Myr$ during which the compressive modes are a collection of discrete episodes. Putting this together, a proxy for the distribution of time spent in compressive tidal modes might consist in taking the \emph{lus} distribution over the time interval $\Delta T_1$, followed by the \emph{tt} statistics over the interval $\Delta T_2$. Fig.~\ref{fig:age} plots the \emph{lus} and \emph{tt} distributions (red and blue lines) showing the usual double exponential, together with the linear combination of the two described above (black curve).

If we take the bold step of identifying the duration of the modes (x-axis on Fig.~\ref{fig:age}), with the age of stellar associations, then the vertical y-axis may be interpreted as the number of associations per age interval, $dN/d\tau$. The black curve on Fig.~\ref{fig:age} can be fitted with a $\tau^{-1.08}$ power law ($\sigma = 2\times 10^{-4}$) over the interval $t \in [10, 100] \Myr$). It is striking that the duration of the compressive modes is very similar to the $\tau^{-1}$ distribution derived from HST data by \citet{Fall2005} for the age of young star clusters. While this does not constitute a full cause-to-effect chain of reasoning, the unexpected agreement does offer a strong hint that time-dependent tidal fields bear on the demographics of young clusters in galaxy mergers.

\subsection{Energy of compressive modes}
\begin{figure}
\plotone{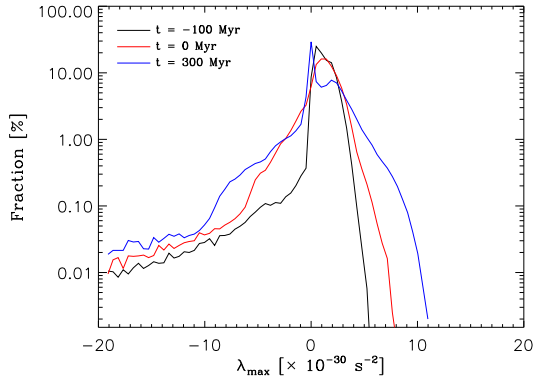}
\caption{Normalized distribution (in percent) of the maximum eigenvalue ($\lambda_\mathrm{max}$) of the tidal tensor for the Antennae run. Three key steps are shown: the configuration in isolation ($t = -100 \Myr$), the first pericenter passage ($t = 0 \Myr$) and the merger stage ($t = 300 \Myr$).}
\label{fig:distrib}
\end{figure}

To characterize statistically the strength of the tides, we plot in Fig~\ref{fig:distrib} the distribution of the maximum eigenvalue of the tidal tensor for all particles, at selected times. As mentioned before, a negative value stands for a fully compressive mode. It is clear that, at any stage of the merger, the distribution can not be fitted by simple analytical functions (such as a Gaussian or power laws). Its high degree of asymmetry with respect to the origin mainly stems from the distribution of eigenvalues in the galaxies in isolation (black curve on Fig.~\ref{fig:distrib}), where a large fraction of the mass ($\sim 97\%$) experiences extensive but weak tidal modes, while the few percent mass elements nearest to the galactic center experience compressive modes of large, negative eigenvalues. The distribution of compressive and extensive modes tends to become more even-handed in the course of the merge, in terms of the occupation percentage at given $|\lambda|$. This is illustrated at the first pericenter passage, $t = 0 \Myr$, and at $t = 300 \Myr$ when the merger is in full swing. One can deduce from this that the strong destructive effect of the extensive tidal field acting at some point in the system is balanced out, on average, by a compressive mode of comparable strength in some other region, like e.g. the tidal tails. Given that individual orbits enter and exit compressive modes several times during the merger (cf. Fig.~\ref{fig:orbits_particles}) one may anticipate that transitions from compressive to extensive modes are relatively sharp over a timescale of $\sim 2 \Myr$.

We already noted how the full impact of the compressive tides could only be quantified once the hydrodynamics of star formation is included in the modeling. Despite this caveat, one can estimate the relative importance of compressive tides by comparing the extra binding energy they add to a cluster, to the feedback from stellar winds. For simplicity, let us assume an isotropic tidal field, i.e. described by a tensor set equal to the identity matrix times $\lambda$. For a mass distribution of total mass $M_\mathrm{t}$, the corresponding binding energy driven by the tide is
\begin{equation}
\label{eqn:energy}
E_\mathrm{T, cluster} = -\frac{1}{2} \lambda \int_0^{M_\mathrm{t}}{r^2 \ dm} = -\frac{1}{2} \lambda \alpha \ M_\mathrm{t} R_\mathrm{t}^2
\end{equation}
where $\alpha$ is a dimensionless quantity of the order of unity, which encapsulates details of the mass distribution, and $R_\mathrm{t}$ is the truncation radius. Applying Eq.~\ref{eqn:energy} to the resolution of our simulations ($M_\mathrm{t} = 10^5 \msun$ and $R_\mathrm{t} = 10^2 \U{pc}$), we get a typical value for $\lambda$ of $\sim -10^{-30} \U{s^{-2}}$ ($< 0 $ for a compressive tide). This yields an estimate for the tidal binding energy of $\sim 10^{49} \U{ergs}$. Clearly, this value does not compare to the energy release $\sim 10^{51} \U{ergs}$ of a single supernova explosion, and hence SN-driven gas expulsion would proceed largely unaffected by the tides. However, the binding energy computed from Eq.~\ref{eqn:energy} would compare favorably to the kinetic energy of an O-star's wind \citep{Cappa1998, Martin2008}. In that case the mass loss driven by the winds of early-type stars would be slowed down considerably. The enhanced mass retention would then help stop or reduce the dissolution of young clusters, an aspect that has immediate consequences to the question related to their mass functions in mergers.

In addition to star clusters, compressive tidal modes might also be important for the formation of TDGs. Detailed high-resolution $N$-body simulations showed that the formation of TDGs is virtually impossible in purely stellar systems \citep{Wetzstein2007}. Consequently, a dissipative component (gas) is a necessary ingredient to form TDGs. Because the compressive region will span kiloparsec-size volumes, enough stars may form that a TDG will ensue. The binding energy of the tidal field estimated from Eq.~\ref{eqn:energy} for a mass $M_\mathrm{t} = 10^9 \msun$ and a size of $10 \kpc$ (typical values for the compressive regions found in the outer parts of our merger simulations) is $\sim 10^{57} \U{ergs}$, or the kinetic energy release of $10^6$ type-II supernovae. Now, if one converts $10^9 \msun$ of gas using a standard stellar initial mass function and a star formation efficiency (SFE) of $10-20 \%$, one ends up with on order of a million type-II supernova candidate stars. While this calculation is rough, it does allow us to conclude that the energy reservoir of compressive tides must be factored in more detailed analysis of TDG formation models, as it is at least comparable to SN energy feedback.

\subsection{Perspectives}
In this last, more speculative section we explore briefly three key areas for future work that will draw from the current study:

\begin{itemize}
\item The first and more obvious is the modeling of the time-evolution of cluster mass functions driven by galactic tides. A number of previous studies derived cluster dissolution rates based on steady external tides \citep{Baumgardt2003, Gieles2008a, Gieles2008b, Kuepper2008, Vesperini2009, Ernst2009}. Other approaches encapsulate the tide's time-evolution in an approximate ``shock'' treatment as the model cluster orbits the host galaxy (\citealt{Spitzer1987}; see \citealt{Gieles2007} for a recent application to spiral galaxies). We note that the tidal tensor approach taken in this paper encompasses those mentioned here as limiting cases. But the important point to stress in our view is that the statistics of tidal fields change very significantly once we consider out-of-equilibrium dynamics. It will be important in future to take into account this evolution of the tidal field itself when considering the evolution of cluster mass distributions, especially in mergers.

\item Secondly, the tidal field bears down on star-forming regions  at the mass- and length scales of star clusters in merging galaxies, and hence on the hydrodynamics of fragmentation and star formation in such systems. Future modeling of the first $\sim 10^6 \U{yr}$ of a giant molecular cloud embedded in a realistic, fully evolving background tidal field should shed new light on questions pertaining to the survival of young (open or globular) clusters \citep{Bastian2006, Goodwin2008}. The survival of clusters is set from a combination of several effects, e.g. the expulsion of the residual gas \citep{Goodwin1997, Geyer2001, Chen2008}; and two-body relaxation \citep{Portegies2008}. How the rapid evolution of the tides that we have reported here changes the picture derived from the secular evolution of clusters (the weak field limit, see e.g. \citealt{Goodwin1997, Boily2003a, Boily2003b, Baumgardt2007, Parmentier2009}) is an aspect worthy of further consideration.

\item Finally, several authors have confirmed  in recent years the presence of multiple stellar main sequences (MS) in resolved massive clusters \citep[see e.g][]{Piotto2007, Milone2009}. \citet{dAntona2004} suggested that a stellar initial mass function (IMF) with a flat slope at the high-mass end would enhance CNO abundances yet leave relatively unchanged the abundances of iron-peak elements, providing a fit for the multiple threads in cluster HR diagrams. It is unclear why a bias in the stellar IMF should develop in some clusters. One possibility comes in the form of an age-spread as broad as $\sim 10 \Myr$, which could be interpreted as a second episode of star formation. This would open up the possibility that (i) clusters are self-enriched (by retaining slow stellar winds); or (ii) accrete gas from gas-rich environments met on their galactic orbit. In that context it is clear that compressive gravitational tides would help increase the likelihood of a second episode of star formation. Gas accretion by the cluster should impact on \emph{both} the CNO abundances and those of the iron-peak elements, and so offers little hope of a viable explanation to this problem. Furthermore, the cluster has little time over $10 \Myr$ to move significantly away from the volume where it first condensed before the second burst of star formation takes place. Thus, this scenario would preserve the homogeneity of the metal abundances the cluster already holds. Self-enrichment, on the other hand, should be more effective in broadening CNO abundances since the new abundances would remain correlated with those of the host cluster at birth. Recent modeling of rotating massive stars has shown that slow disk-like ejecta may be more effectively retained by the cluster than was assumed up to now \citep{dErcole2008, Decressin2007}. Clearly the retention of light elements would be enhanced if the cluster was undergoing a compressive tidal mode at the time when such winds were operative. It is too early to pin down quantitatively the actual impact of tidal modes on such cluster evolutionary processes, which will surely require exhaustive modeling of the hydrodynamics and the chemical evolution, not done here.
\end{itemize}

\acknowledgments
We thank the referee who provided interesting comments that resulted in substantial improvement of this paper. FR is a member of the IK~I033-N \emph{Cosmic Matter Circuit} at the University of Vienna. TN acknowledges support from the DFG cluster of excellence `Origin and Structure of the Universe'. This project has been supported by the DFG Priority Program 1177 `Galaxy Evolution'.


\begin{thebibliography}{}
\bibitem[Barnes(1988)]{Barnes1988} Barnes, J.~E.\ 1988, \apj, 331, 699
\bibitem[Barnes(2004)]{Barnes2004} Barnes, J.~E.\ 2004, \mnras, 350, 798
\bibitem[Barton, Geller, \& Kenyon(2000)]{Barton2000} Barton, E.~J., Geller, M.~J., \& Kenyon, S.~J.\ 2000, \apj, 530, 660
\bibitem[Bastian \& Goodwin(2006)]{Bastian2006} Bastian, N., \& Goodwin, S.~P.\ 2006, \mnras, 369, L9
\bibitem[Baumgardt \& Makino(2003)]{Baumgardt2003} Baumgardt, H., \& Makino, J.\ 2003, \mnras, 340, 227
\bibitem[Baumgardt \& Kroupa(2007)]{Baumgardt2007} Baumgardt, H., \& Kroupa, P.\ 2007, \mnras, 380, 1589
\bibitem[Bekki \& Chiba(2002)]{Bekki2002a} Bekki, K., \& Chiba, M.\ 2002, \apj, 566, 245
\bibitem[Bekki et al.(2002)]{Bekki2002b} Bekki, K., Forbes, D.~A., Beasley, M.~A., \& Couch, W.~J.\ 2002, \mnras, 335, 1176
\bibitem[Boily et al.(2001)]{Boily2001} Boily, C.~M., Kroupa, P., \& Pe{\~n}arrubia-Garrido, J.\ 2001, New Astronomy, 6, 27
\bibitem[Boily \& Kroupa(2003a)]{Boily2003a} Boily, C.~M., \& Kroupa, P.\ 2003a, \mnras, 338, 665
\bibitem[Boily \& Kroupa(2003b)]{Boily2003b} Boily, C.~M., \& Kroupa, P.\ 2003b, \mnras, 338, 673
\bibitem[Bonnor(1956)]{Bonnor1956} Bonnor, W.~B.\ 1956, \mnras, 116, 351
\bibitem[Bournaud et al.(2005)]{Bournaud2005} Bournaud, F., Jog, C.~J., \& Combes, F.\ 2005, \aap, 437, 69
\bibitem[Bournaud et al.(2008)]{Bournaud2008} Bournaud, F., Duc, P.-A., \& Emsellem, E.\ 2008, \mnras, 389, L8
\bibitem[Bridge et al.(2007)]{Bridge2007} Bridge, C.~R., et al.\ 2007, \apj, 659, 931
\bibitem[Cappa \& Benaglia(1998)]{Cappa1998} Cappa, C.~E., \& Benaglia, P.\ 1998, \aj, 116, 1906
\bibitem[Chen \& Ko(2008)]{Chen2008} Chen, H.-C., \& Ko, C.-M.\ 2008, Astronomische Nachrichten, 329, 1053
\bibitem[Decressin et al.(2007)]{Decressin2007} Decressin, T., Meynet, G., Charbonnel, C., Prantzos, N., \& Ekstr{\"o}m, S.\ 2007, \aap, 464, 1029
\bibitem[de Grijs \& Parmentier(2007)]{deGrijs2007} de Grijs, R., \& Parmentier, G.\ 2007, Chinese Journal of Astronomy and Astrophysics, 7, 155
\bibitem[Dehnen(1993)]{Dehnen1993} Dehnen, W.\ 1993, \mnras, 265, 250
\bibitem[Dehnen(2001)]{Dehnen2001} Dehnen, W.\ 2001, \mnras, 324, 273
\bibitem[Dehnen(2002)]{Dehnen2002} Dehnen, W.\ 2002, Journal of Computational Physics, 179, 27
\bibitem[di Matteo et al.(2007)]{diMatteo2007} di Matteo, P., Combes, F., Melchior, A.-L., \& Semelin, B.\ 2007, \aap, 468, 61
\bibitem[di Matteo et al.(2008)]{diMatteo2008} di Matteo, P., Bournaud, F., Martig, M., Combes, F., Melchior, A.-L., \& Semelin, B.\ 2008, \aap, 492, 31
\bibitem[Dekel et al.(2003)]{Dekel2003} Dekel, A., Devor, J., \& Hetzroni, G.\ 2003, \mnras, 341, 326
\bibitem[D'Antona \& Caloi(2004)]{dAntona2004} D'Antona, F., \& Caloi, V.\ 2004, \apj, 611, 871
\bibitem[D'Ercole et al.(2008)]{dErcole2008} D'Ercole, A., Vesperini, E., D'Antona, F., McMillan, S.~L.~W., \& Recchi, S.\ 2008, \mnras, 391, 825
\bibitem[Dobbs \& Pringle(2009)]{Dobbs2009} Dobbs, C.~L., \& Pringle, J.~E.\ 2009, \mnras, 665
\bibitem[Ebert(1955)]{Ebert1955} Ebert, R.\ 1955, Zeitschrift fur Astrophysik, 37, 217
\bibitem[Elmegreen(1989)]{Elmegreen1989} Elmegreen, B.~G.\ 1989, \apj, 338, 178
\bibitem[Elmegreen \& Efremov(1997)]{Elmegreen1997} Elmegreen, B.~G., \& Efremov, Y.~N.\ 1997, \apj, 480, 235
\bibitem[Elmegreen et al.(2007)]{Elmegreen2007} Elmegreen, D.~M., Elmegreen, B.~G., Ferguson, T., \& Mullan, B.\ 2007, \apj, 663, 734
\bibitem[Ernst, Just, \& Spurzem(2009)]{Ernst2009} Ernst, A., Just, A., \& Spurzem, R.\ 2009, \mnras, 1173
\bibitem[Fall et al.(2005)]{Fall2005} Fall, S.~M., Chandar, R., \& Whitmore, B.~C.\ 2005, \apjl, 631, L133
\bibitem[Geyer \& Burkert(2001)]{Geyer2001} Geyer, M.~P., \& Burkert, A.\ 2001, \mnras, 323, 988
\bibitem[Gieles et al.(2007)]{Gieles2007} Gieles, M., Athanassoula, E., \& Portegies Zwart, S.~F.\ 2007, \mnras, 376, 809
\bibitem[Gieles \& Baumgardt(2008)]{Gieles2008a} Gieles, M., \& Baumgardt, H.\ 2008, \mnras, 389, L28
\bibitem[Gieles \& Bastian(2008)]{Gieles2008b} Gieles, M., \& Bastian, N.\ 2008, \aap, 482, 165
\bibitem[Gieles(2009)]{Gieles2009} Gieles, M.\ 2009, \mnras, 277
\bibitem[Gilbert \& Graham(2007)]{Gilbert2007} Gilbert, A.~M., \& Graham, J.~R.\ 2007, \apj, 668, 168
\bibitem[Gingold \& Monaghan(1977)]{Gingold1977} Gingold, R.~A., \& Monaghan, J.~J.\ 1977, \mnras, 181, 375
\bibitem[Goodwin(1997)]{Goodwin1997} Goodwin, S.~P.\ 1997, \mnras, 284, 785
\bibitem[Goodwin(2008)]{Goodwin2008} Goodwin, S.~P 2008, arXiv:0802.2207
\bibitem[Grebel \& Brandner(2002)]{Grebel2002} Grebel, E.K. \& Brandner, W. (eds)\ 2002, Modes of Star Formation and the Origin of Field Populations, PASP Conf. Series Vol. 285, (San Francisco: ASP), 457 pages
\bibitem[Harris \& Pudritz(1994)]{Harris1994} Harris, W.~E., \& Pudritz, R.~E.\ 1994, \apj, 429, 177
\bibitem[Hernquist(1990)]{Hernquist1990} Hernquist, L.\ 1990, \apj, 356, 359
\bibitem[Hibbard \& van Gorkom(1996)]{Hibbard1996} Hibbard, J.~E., \& van Gorkom, J.~H.\ 1996, \aj, 111, 655
\bibitem[Hibbard et al.(2001)]{Hibbard2001} Hibbard, J.~E., van der Hulst, J.~M., Barnes, J.~E., \& Rich, R.~M.\ 2001, \aj, 122, 2969
\bibitem[Hibbard et al.(2005)]{Hibbard2005} Hibbard, J.~E., et al.\ 2005, \apjl, 619, L87
\bibitem[Johansson, Naab, \& Burkert(2009)]{Johansson2009} Johansson, P.~H., Naab, T., \& Burkert, A.\ 2009, \apj, 690, 802
\bibitem[Karl et al.(2008)]{Karl2008} Karl, S.~J., Naab, T., Johansson, P.~H., Theis, C., \& Boily, C.~M.\ 2008, Astronomische Nachrichten, 329, 1042
\bibitem[Kim, Wise, \& Abel(2009)]{Kim2009} Kim, J.-h., Wise, J.~H., \& Abel, T.\ 2009, \apjl, 694, L123
\bibitem[Klessen, Burkert, \& Bate(1998)]{Klessen1998} Klessen, R.~S., Burkert, A., \& Bate, M.~R.\ 1998, \apjl, 501, L205
\bibitem[Kopp(2008)]{Kopp2008} Kopp, J.\ 2008, Int. J. Mod. Phys., C19, 523
\bibitem[Kuijken \& Dubinski(1995)]{Kuijken1995} Kuijken, K., \& Dubinski, J.\ 1995, \mnras, 277, 1341
\bibitem[Kumai \& Tosa(1986)]{Kumai1986} Kumai, Y., \& Tosa, M.\ 1986, \apss, 119, 211
\bibitem[K{\"u}pper et al.(2008)]{Kuepper2008} K{\"u}pper, A.~H.~W., MacLeod, A., \& Heggie, D.~C.\ 2008, \mnras, 387, 1248
\bibitem[Laine et al.(2003)]{Laine2003} Laine, S., van der Marel, R.~P., Rossa, J., Hibbard, J.~E., Mihos, J.~C., B{\"o}ker, T., \& Zabludoff, A.~I.\ 2003, \aj, 126, 2717 
\bibitem[Li et al.(2008)]{Li2008} Li, C., Kauffmann, G., Heckman, T.~M., Jing, Y.~P., \& White, S.~D.~M.\ 2008, \mnras, 385, 1903
\bibitem[Li et al.(2004)]{Li2004} Li, Y., Mac Low, M.-M., \& Klessen, R.\ 2004, Bulletin of the American Astronomical Society, 36, 1482 
\bibitem[Mac Low \& Klessen(2004)]{MacLow2004} Mac Low, M.-M., \& Klessen, R.~S.\ 2004, Reviews of Modern Physics, 76, 12
\bibitem[Madau et al.(1996)]{Madau1996} Madau, P., Ferguson, H.~C., Dickinson, M.~E., Giavalisco, M., Steidel, C.~C., \& Fruchter, A.\ 1996, \mnras, 283, 1388
\bibitem[Mart{\'{\i}}n et al.(2008)]{Martin2008} Mart{\'{\i}}n, M.~C., Cappa, C.~E., \& Benaglia, P.\ 2008, Revista Mexicana de Astronomia y Astrofisica Conference Series, 33, 164
\bibitem[McMillan \& Dehnen(2007)]{McMillan2007} McMillan, P.~J., \& Dehnen, W.\ 2007, \mnras, 378, 541
\bibitem[Mengel et al.(2002)]{Mengel2002} Mengel, S., Lehnert, M.~D., Thatte, N., \& Genzel, R.\ 2002, Extragalactic Star Clusters, 207, 378
\bibitem[Mengel et al.(2005)]{Mengel2005} Mengel, S., Lehnert, M.~D., Thatte, N., \& Genzel, R.\ 2005, \aap, 443, 41
\bibitem[Merritt(1996)]{Merritt1996} Merritt, D.\ 1996, \aj, 111, 2462
\bibitem[Meurer(1995)]{Meurer1995} Meurer, G.~R.\ 1995, \nat, 375, 742
\bibitem[Mihos \& Hernquist(1996)]{Mihos1996} Mihos, J.~C., \& Hernquist, L.\ 1996, \apj, 464, 641
\bibitem[Milone et al.(2009)]{Milone2009} Milone, A.~P., Stetson, P.~B., Piotto, G., Bedin, L.~R., Anderson, J., Cassisi, S., \& Salaris, M.\ 2009, \aap, 503, 755
\bibitem[Mirabel et al.(1992)]{Mirabel1992} Mirabel, I.~F., Dottori, H., \& Lutz, D.\ 1992, \aap, 256, L19
\bibitem[Naab \& Burkert(2003)]{Naab2003} Naab, T., \& Burkert, A.\ 2003, \apj, 597, 893
\bibitem[Naab, Jesseit, \& Burkert(2006)]{Naab2006} Naab, T., Jesseit, R., \& Burkert, A.\ 2006, \mnras, 372, 839
\bibitem[Navarro et al.(1997)]{Navarro1997} Navarro, J.~F., Frenk, C.~S., \& White, S.~D.~M.\ 1997, \apj, 490, 493
\bibitem[Parmentier \& Fritze(2009)]{Parmentier2009} Parmentier, G., \& Fritze, U.\ 2009, \apj, 690, 1112
\bibitem[Piotto et al.(2007)]{Piotto2007} Piotto, G., et al.\ 2007, \apjl, 661, L53
\bibitem[Plummer(1911)]{Plummer1911} Plummer, H.~C.\ 1911, \mnras, 71, 460
\bibitem[Portegies Zwart \& Chen(2008)]{Portegies2008} Portegies Zwart, S.~F., \& Chen, H.-C.\ 2008, Mass Loss from Stars and the Evolution of Stellar Clusters, 388, 329
\bibitem[Renaud et al.(2008)]{Renaud2008} Renaud, F., Boily, C.~M., Fleck, J.-J., Naab, T., \& Theis, C.\ 2008, \mnras, 391, L98
\bibitem[Robertson \& Kravtsov(2008)]{Robertson2008} Robertson, B.~E., \& Kravtsov, A.~V.\ 2008, \apj, 680, 1083
\bibitem[Sandage(1961)]{Sandage1961} Sandage, A.\ 1961, Washington: Carnegie Institution, 1961
\bibitem[Schweizer(1978)]{Schweizer1978} Schweizer, F.\ 1978, Structure and Properties of Nearby Galaxies, 77, 279
\bibitem[Spitzer(1987)]{Spitzer1987} Spitzer, L.\ 1987, Princeton, NJ, Princeton University Press, 1987, 191 p.
\bibitem[Springel \& Hernquist(2003)]{Springel2003} Springel, V., \& Hernquist, L.\ 2003, \mnras, 339, 289
\bibitem[Stewart(2009)]{Stewart2009} Stewart, K.~R.\ 2009, arXiv:0902.2214
\bibitem[Tan, Krumholz, \& McKee(2006)]{Tan2006} Tan, J.~C., Krumholz, M.~R., \& McKee, C.~F.\ 2006, \apjl, 641, L121
\bibitem[Toomre \& Toomre(1972)]{Toomre1972} Toomre, A., \& Toomre, J.\ 1972, \apj, 178, 623
\bibitem[Toomre(1977)]{Toomre1977} Toomre, A.\ 1977, Evolution of Galaxies and Stellar Populations, 401
\bibitem[Valluri(1993)]{Valluri1993} Valluri, M.\ 1993, \apj, 408, 57
\bibitem[Vesperini(1997)]{Vesperini1997} Vesperini, E.\ 1997, \mnras, 287, 915
\bibitem[Vesperini, McMillan, \& Portegies Zwart(2009)]{Vesperini2009} Vesperini, E., McMillan, S.~L.~W., \& Portegies Zwart, S.\ 2009, arXiv:0904.3934
\bibitem[Wetzstein, Naab, \& Burkert(2007)]{Wetzstein2007} Wetzstein, M., Naab, T., \& Burkert, A.\ 2007, \mnras, 375, 805
\bibitem[Whitmore \& Schweizer(1995)]{Whitmore1995} Whitmore, B.~C., \& Schweizer, F.\ 1995, \aj, 109, 960 
\bibitem[Whitmore et al.(2007)]{Whitmore2007} Whitmore, B.~C., Chandar, R., \& Fall, S.~M.\ 2007, \aj, 133, 1067
\end{thebibliography}
\end{document}